\documentclass[fdp,a4paper,fleqn]{w-art}
\usepackage{times,cite,w-thm}
\theoremstyle{plain}

\theoremstyle{definition}

\usepackage{graphicx}
\usepackage{amsmath}
\usepackage{amssymb}
\usepackage{mathrsfs}
\usepackage{dsfont}
\usepackage{epsfig}
\usepackage{graphicx}
\usepackage{subfigure}
\usepackage{cancel}

\newcommand{\om}{\omega}

\def\cC{{\cal C}}

 \def\cO{{\cal O}}

\newcommand{\Gev}{\text{GeV}}

\newcommand{\be}{\begin{equation}}
\newcommand{\ee}{\end{equation}}
\newcommand{\beq}{\begin{equation}}
\newcommand{\eeq}{\end{equation}}
\newcommand{\bac}{\beq\begin{array}}
\newcommand{\eac}{\end{array}\eeq}
\newcommand{\ba}{\begin{array}}
\newcommand{\ea}{\end{array}}
\newcommand{\bea}{\begin{eqnarray}}
\newcommand{\eea}{\end{eqnarray}}
\newcommand{\beaa}{\begin{eqnarray*}}
\newcommand{\eeaa}{\end{eqnarray*}}

\def\dd{\displaystyle}
\newcommand{\nn}{\nonumber}

\begin{document}
\DOIsuffix{theDOIsuffix}
\Volume{55}
\Month{01}
\Year{2007}
\pagespan{1}{}
\Receiveddate{XXXX}
\Reviseddate{XXXX}
\Accepteddate{XXXX}
\Dateposted{XXXX}

\title[TB Mixing and Discrete Symmetries]{Tri-Bimaximal Neutrino Mixing and Discrete Flavour Symmetries}
\author[G. Altarelli]{Guido Altarelli\inst{1,2}\footnote{E-mail:~\textsf{guido.altarelli@cern.ch}}}
\address[\inst{1}]{Dipartimento di Fisica `E.~Amaldi', Universit\`a di Roma Tre, INFN, Sezione di Roma Tre, I-00146 Rome, Italy. Preprint: RM3-TH/12-8}
\address[\inst{2}]{CERN, Department of Physics, Theory Division, CH-1211 Geneva 23, Switzerland. Preprint: CERN-PH-TH/2012-138}
\author[F. Feruglio]{Ferruccio Feruglio\inst{3}\footnote{E-mail:~\textsf{feruglio@pd.infn.it}}}
\address[\inst{3}]{Dipartimento di Fisica `G.~Galilei', Universit\`a di Padova, INFN, Sezione di Padova, Via Marzolo~8, I-35131 Padua, Italy. Preprint: DFPD-2012/TH/5}
\author[L. Merlo]{Luca Merlo\inst{4,5}\footnote{Corresponding author\quad E-mail:~\textsf{luca.merlo@ph.tum.de}}}
\address[\inst{4}]{Physik-Department, Technische Universit\"at M\"unchen, James-Franck-Strasse, D-85748 Garching, Germany. Preprint: TUM-HEP-835/12}
\address[\inst{5}]{Institute for Advanced Study, Technische Universit\"at M\"unchen, Lichtenbergstrasse 2a, D-85748 Garching, Germany}

\begin{abstract}
We review the application of non-Abelian discrete groups to Tri-Bimaximal (TB) neutrino mixing, which is supported by experiment as a possible good first approximation to the data. After summarizing the motivation and the formalism, we discuss specific models, mainly those based on $A_4$ but also on other finite groups, and their phenomenological implications, including the extension to quarks. The recent measurements of $\theta_{13}$ favour versions of these models where a suitable mechanism leads to corrections to $\theta_{13}$ that can naturally be larger than those to $\theta_{12}$ and $\theta_{23}$. The virtues and the problems of TB mixing models are discussed, also in connection with lepton flavour violating processes, and the different approaches are compared.
\end{abstract}

\maketitle

%
%%%%%%%%%%%%%%%%%%%%%%%% 1.  INTRODUCTION   %%%%%%%%%%%%%%%%%%%%%%%%%%%%%%
\section{Introduction}

Neutrino mixing \cite{Altarelli:2004za,Mohapatra:2006gs,Grimus:2006nb,GonzalezGarcia:2007ib,Altarelli:2009wt,Altarelli:2010fk} is important because it could in principle provide new clues for the understanding of the flavour problem. Even more so since neutrino mixing angles show a pattern that is completely different than that of quark mixing. The bulk of the data on neutrino oscillations are well described in terms of three active neutrinos. By now all three mixing angles have been measured,  although with different levels of accuracy (see Tab.~\ref{tab:data} \cite{Fogli:2012ua,Tortola:2012te}). In particular, we have experimental evidence for a non vanishing value of the smallest angle $\theta_{13}$ (see Tab.~\ref{tab:t13} \cite{Abe:2011sj,Adamson:2011qu,Abe:2011fz,An:2012eh}): considering the most precise results from DOUBLE CHOOZ, Daya Bay and RENO, we get
\be
\sin^2\theta_{13}=0.0253\pm0.0035\,,\\
\label{ourTheta13}
\ee
for both the mass orderings.

\begin{vchtable}[h]
\vchcaption{Recent fits to neutrino oscillation data from \cite{Fogli:2012ua,Tortola:2012te}. In the brackets the IH case. $)^\star$ In this case the full $(0,\,2\pi)$ is allowed.}
\label{tab:data}
\begin{tabular}{@{}ccc@{}}
  \hline
  &&\\[-3mm]
  Quantity & Fogli {\it et al.} \cite{Fogli:2012ua} & Schwetz {\it et al.} \cite{Tortola:2012te} \\[1mm]
  \hline
  &&\\[-3mm]
  $\Delta m^2_{sun}~(10^{-5}~{\rm eV}^2)$ 			&$7.54^{+0.26}_{-0.22}$ 																																								&$7.62\pm0.19$  \\[1mm]
  $\Delta m^2_{atm}~(10^{-3}~{\rm eV}^2)$ 			&$2.43^{+0.07}_{-0.09}$ ($2.42^{+0.07}_{-0.1}$) 																															&$2.53^{+0.08}_{-0.10}$ ($2.40^{+0.10}_{-0.07}$)  \\[1mm]
  $\sin^2\theta_{12}$ 											&$0.307^{+0.018}_{-0.016}$ 																																							&$0.320^{+0.015}_{-0.017}$ \\[1mm]
  $\sin^2\theta_{23}$ 											&$0.398^{+0.03}_{-0.026}$ ($0408^{+0.035}_{-0.03}$) 																													&$0.49^{+0.08}_{-0.05}$ ($0.53^{+0.05}_{-0.07}$) \\[1mm]
  $\sin^2\theta_{13}$ 											&$0.0245^{+0.0034}_{-0.0031}$ ($0.0246^{+0.0034}_{-0.0031}$) 																									&$0.026^{+0.003}_{-0.004}$ ($0.027^{+0.003}_{-0.004}$)  \\[1mm]
  $\delta_{CP}/\pi$ 												&$0.89^{+0.29}_{-0.44}$ ($0.90^{+0.32}_{-0.43}$)																															&$0.83^{+0.54}_{-0.64}$ ($0.07^\star$)\\
  \hline
\end{tabular}
\end{vchtable}

\begin{vchtable}[h!]
\vchcaption{The reactor angle measurements from the recent experiments T2K\cite{Abe:2011sj}, MINOS\cite{Adamson:2011qu}, DOUBLE CHOOZ\cite{Abe:2011fz}, Daya Bay \cite{An:2012eh} and RENO \cite{Ahn:2012nd}, for the normal (inverse) hierarchy.}
\label{tab:t13}
\begin{tabular}{@{}ccc@{}}
  \hline
  &&\\[-3mm]
  Quantity 	& $\sin^22\theta_{13}$ & $\sin^2\theta_{13}$ \\[1mm]
  \hline
  &&\\[-2mm]
  T2K\cite{Abe:2011sj} 					& 	$0.11^{+0.11}_{-0.05}$ ($0.14^{+0.12}_{-0.06}$)				
  													& 	$0.028^{+0.019}_{-0.024}$ ($0.036^{+0.022}_{-0.030}$) \\[1mm]
  MINOS\cite{Adamson:2011qu} 	&	$0.041^{+0.047}_{-0.031}$ ($0.079^{+0.071}_{-0.053}$)		
  													& 	$0.010^{+0.012}_{-0.008}$ ($0.020^{+0.019}_{-0.014}$) \\[1mm]
  DC\cite{Abe:2011fz} 					&	$0.086\pm0.041\pm0.030$ 				& $0.022^{+0.019}_{-0.018}$ 			\\[1mm]
  DYB\cite{An:2012eh}					&	$0.092\pm0.016\pm0.005$				& $0.024\pm0.005	$ 							\\[1mm]
  RENO\cite{Ahn:2012nd}				&	$0.113\pm0.013\pm0.019$				& $0.029\pm0.006	$ 			\\[1mm]
  \hline
\end{tabular}
\end{vchtable}

Models of neutrino mixing based on discrete flavour groups have received a lot of attention in recent years \cite{Altarelli:2010gt,Ishimori:2010au,Ludl:2010bj,Grimus:2010ak,Parattu:2010cy,Grimus:2011fk}. There are a number of special mixing patterns that have been studied in that context. Most of these mixing matrices have $\sin^2{\theta_{23}}=1/2$, $\sin^2{\theta_{13}}=0$,
values that are a good approximation to the data, and differ by the value of the solar angle $\sin^2{\theta_{12}}$.  The observed $\sin^2{\theta_{12}}$, the best measured mixing angle,  is very close, from below, to the so called Tri-Bimaximal (TB) value \cite{Harrison:2002er,Harrison:2002kp,Xing:2002sw,Harrison:2002et,Harrison:2003aw} which is $\sin^2{\theta_{12}}=1/3$   (see Fig.~\ref{fig:TBGRBM}). Alternatively it is also very close, from above, to the Golden Ratio (GR) value \cite{Kajiyama:2007gx,Everett:2008et,Ding:2011cm,Feruglio:2011qq} which is $\sin^2{\theta_{12}}=\frac{1}{\sqrt{5}\phi} = \frac{2}{5+\sqrt{5}}\sim 0.276$, where $\phi= (1+\sqrt{5})/2$ is the GR (for a different connection to the GR, see Refs.~\cite{Rodejohann:2008ir,Adulpravitchai:2009bg}). On a different perspective, one has considered models with Bi-Maximal (BM) mixing, with $\sin^2{\theta_{12}}=1/2$, i.e. also maximal, as the neutrino mixing matrix before diagonalization of charged leptons. This is in line with the well-known empirical observation that $\theta_{12}+\theta_C\sim \pi/4$, where $\theta_C$ is the Cabibbo angle, a relation known as quark-lepton complementarity \cite{Altarelli:2004jb,Raidal:2004iw,Minakata:2004xt,Frampton:2004vw,Ferrandis:2004vp,Kang:2005as,Li:2005ir,Cheung:2005gq,Xing:2005ur,Datta:2005ci,Antusch:2005ca,Lindner:2005pk,Minakata:2005rf,Ohlsson:2005js,King:2005bj,Dighe:2006zk,Chauhan:2006im,Hochmuth:2006xn,Schmidt:2006rb,Plentinger:2006nb,Plentinger:2007px}. Probably the exact complementarity relation becomes more plausible if replaced with $\theta_{12}+\mathcal{O}(\theta_C)\sim \pi/4$ (which we call ``weak'' complementarity). One can think of models where, because of a suitable symmetry,  BM mixing holds in the neutrino sector at leading order and the necessary, rather large, corrective terms for $\theta_{12}$ arise from the diagonalization of the charged lepton mass matrices \cite{Raidal:2004iw,Minakata:2004xt,Minakata:2005rf,Frampton:2004vw,Ferrandis:2004vp,Kang:2005as,Altarelli:2004jb,Li:2005ir,Cheung:2005gq,Xing:2005ur,Datta:2005ci,Ohlsson:2005js,Antusch:2005ca,Lindner:2005pk,King:2005bj,Dighe:2006zk,Chauhan:2006im,Schmidt:2006rb,Hochmuth:2006xn,Plentinger:2006nb,Plentinger:2007px,Altarelli:2009gn,Toorop:2010yh,Patel:2010hr,Meloni:2011fx,Shimizu:2010pg,Ahn:2011yj}. These coincidences cannot all be relevant and perhaps all of them are pure accidents. But if one or the other of these coincidences is taken seriously then one is led to consider models where TB or GR or BM mixing are naturally predicted as a good first approximation. 

In the following we will concentrate on TB mixing which is perhaps the most plausible and certainly the most studied first approximation to the data. The simplest symmetry that, in leading order (LO), leads to TB is $A_4$, the group of even permutations of 4 objects, a subgroup of $S_4$ that includes all 4-object permutations. Thus, in the following, we will devote a special attention to $A_4$ models, but alternative theories of TB mixing will also be briefly considered. The plan of the paper is as follows. In Sect.~\ref{sec:NuPatters} we recall the definitions of TB, GR and BM mixing and the symmetries of the corresponding mass matrices. In Sect.~\ref{sec:A4group} we summarize the group theory of $A_4$. In Sect.~\ref{sec:A4Models} we review the structure of $A_4$ models of lepton masses and mixings and, in two separate subsections, we first describe the baseline models and then those special models \cite{Altarelli:2012bn,Lin:2009bw} where additional dynamical ingredients allow that the angle $\theta_{13}$ can naturally be of different (and larger) order of magnitude than the deviations of $\theta_{12}$ from the TB value. We also discuss the comparison with present data of the two options. In Sect.~\ref{sec:Extensions} we discuss the possible extension of the TB models to include quarks, possibly also in  a GUT context. Our speculations on the origin of $A_4$ either as a subgroup of the modular group or as a remnant of an extra dimensional spacetime symmetry are presented in Sect.~\ref{sec:A4origin}. A number of alternative theories of TB mixing are briefly considered in Sect.~\ref{sec:Alternatives}. Sect.~\ref{sec:LFVandLepto} contains a summary on the implications for lepton flavour violation of the different models described in Ref.~\cite{Altarelli:2012bn}. Finally in Sect.~\ref{sec:Conclusions} we derive our conclusions.

\begin{vchfigure}
\includegraphics[width=10.0 cm]{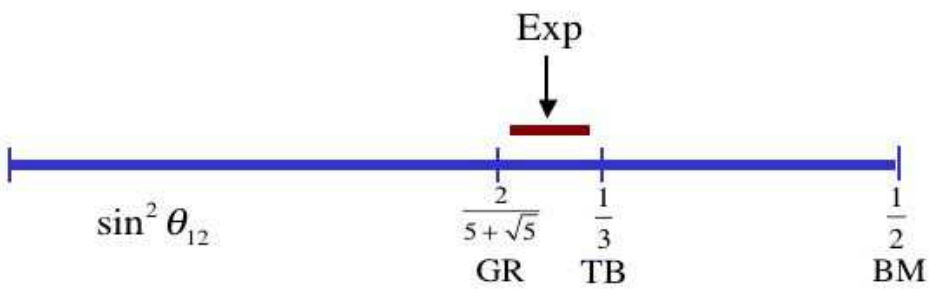}
\vchcaption{The values of $\sin^2{\theta_{12}}$ for TB or GR or BM mixing are compared with the data at $1\sigma$.}
\label{fig:TBGRBM}
\end{vchfigure}

%
%%%%%%%%%%%%%%%%%%%%%%%% 2.  Special Patterns of Neutrino Mixing   %%%%%%%%%%%%%%%%
%
 \section{Special Patterns of Neutrino Mixing}
 \label{sec:NuPatters}

Starting from the PNMS mixing matrix $U$ (we refer the reader to Ref. \cite{Altarelli:2004za, Altarelli:2009wt} for its general definition and parametrisation), the general form of the neutrino mass matrix, in terms of the (complex\footnote{We absorb the Majorana phases in the mass eigenvalues $m_i$, rather than in the mixing matrix $U$.
The dependence on these phases drops in neutrino oscillations.}) mass eigenvalues $m_1, m_2, m_3$, in the basis where charged leptons are diagonal, is given  by
\beq
m_{\nu}=U^* {\rm diag}(m_1,m_2,m_3) U^\dagger\,.
\label{numass}
\eeq
We present here some particularly relevant forms of $U$ and $m_\nu$ that are important in the following. We start by the most general mass matrix that corresponds to $\theta_{13}=0$ and $\theta_{23}$ maximal, that is to $U$ given by  (in a particular phase convention)
\begin{equation}
U= \left(\begin{array}{ccc}
c_{12}					&s_{12}					&0\\
-s_{12}/\sqrt 2		&c_{12}/\sqrt 2		&-1/\sqrt 2\\
-s_{12}/\sqrt 2		&c_{12}/\sqrt 2		&1/\sqrt 2\\
\end{array}\right)\,,
\label{2.1}
\end{equation}
with $c_{12}\equiv \cos{\theta_{12}}$ and $s_{12}\equiv \sin{\theta_{12}}$. 
By applying eq.~(\ref{numass}) we obtain a matrix of the form \cite{Fukuyama:1997ky,Mohapatra:1998ka,Ma:2001mr,Lam:2001fb,Kitabayashi:2002jd,Ghosal:2003mq,Grimus:2003vx,Koide:2003rx,deGouvea:2004gr,Grimus:2004cc,Mohapatra:2004hta,Mohapatra:2005ra,Kitabayashi:2005fc,Mohapatra:2005wk,Mohapatra:2005yu,Ahn:2006nu}:
\begin{equation}
m=\left(\begin{array}{ccc}
x&y&y\\
y&z&w\\
y&w&z
\end{array}\right)\,,
\label{gl}
\end{equation}
with complex coefficients $x$, $y$, $z$ and $w$.
This matrix is the most general one that is symmetric under 2-3 (or $\mu - \tau$) exchange or
\bea
m_\nu=A_{23}m_\nu A_{23}\,,
\label{inv.1}
\eea
where $A_{23}$ is given by
\be
A_{23}=\left(
\begin{array}{ccc}
1&0&0\\
0&0&1\\
0&1&0
\end{array}
\right)\,.
\label{Amutau}
\ee
The solar mixing angle $\theta_{12}$ is given by
\beq
\sin^2 2 \theta_{12}=\,\dfrac{8\vert x^* y+y^*(w+z)\vert^2}{8\vert x^* y+y^*(w+z)\vert^2+(\vert w+z\vert^2-\vert x\vert^2)^2}=\,\dfrac{8 y^2}{(x-w-z)^2+8 y^2}\,,
\label{teta12}
\eeq
where the second equality applies to real parameters. 
Since $\theta_{13}=0$, in this limit there is not no CP violation in neutrino oscillations, and the only physical phases
are the Majorana ones, accounted for by the general case of complex parameters.
We restrict here our consideration to real parameters. There are four of them in eq.~(\ref{gl}) which correspond to the three mass eigenvalues and one remaining mixing angle, $\theta_{12}$. 
Models with $\mu$-$\tau$ symmetry have been extensively studied \cite{Fukuyama:1997ky,Mohapatra:1998ka,Ma:2001mr,Lam:2001fb,Kitabayashi:2002jd,Grimus:2003vx,Koide:2003rx,Ghosal:2003mq,Grimus:2004cc,deGouvea:2004gr,Mohapatra:2005yu,Kitabayashi:2005fc,Mohapatra:2004hta,Mohapatra:2005ra,Mohapatra:2005wk,Ahn:2006nu,Ge:2010js,Hagedorn:2010mq}.

The particularly important case of TB mixing is obtained when $\sin^2{2\theta_{12}}=8/9$ or $x+y=w+z$
\footnote{The other solution $x-y=w+z$ gives rise to TB mixing in another phase convention and is physically equivalent to  $x+y=w+z$.}. In this case the matrix $m_\nu$ takes the form
\begin{equation}
m_\nu=\left(\begin{array}{ccc}
x&y&y\\
y&x+v&y-v\\
y&y-v&x+v
\end{array}\right)\,,
\label{gl21}
\end{equation}
In fact, in this case, $U=U_{TB}$ is given by \cite{Harrison:2002er,Harrison:2002kp,Xing:2002sw,Harrison:2002et,Harrison:2003aw}
\begin{equation}
U_{TB}= \left(\begin{array}{ccc}
\sqrt{2/3}	&1/\sqrt 3	&0\\
-1/\sqrt 6	&1/\sqrt 3	&-1/\sqrt 2\\
-1/\sqrt 6	&1/\sqrt 3	&1/\sqrt 2
\end{array}\right)\,.
\label{2}
\end{equation}
Note that $U_{TB}$ is a rotation matrix with special, fixed angles: indeed all the entries of $U_{TB}$ are pure numbers. This property is related to the particular pattern of $m_{\nu}$ which belongs to the category of the form diagonalizable mass matrices \cite{Low:2003dz}. These matrices are diagonalized by unitary transformations that are independent from the eigenvalues. At the LO discrete flavour models give rise to
form diagonalizable mass matrices and the physical mixing angles are thus unrelated to masses. From eq.~(\ref{numass}), one obtains
\begin{equation}
m_{\nu}=  m_1\,\Phi_1\, \Phi_1^T + m_2\,\Phi_2\, \Phi_2^T + m_3\,\Phi_3\, \Phi_3^T\,, 
\label{1k1}
\end{equation}
where
\be
\Phi_1^T=\frac{1}{\sqrt{6}}(2,-1,-1)\,,\qquad\qquad
\Phi_2^T=\frac{1}{\sqrt{3}}(1,1,1)\,,\qquad\qquad
\Phi_3^T=\frac{1}{\sqrt{2}}(0,-1,1)
\label{4k1}
\ee
are the respective columns of $U_{TB}$ and $m_i$ are the neutrino mass eigenvalues.
It is easy to see that the TB mass matrix in eqs.~(\ref{1k1}) and (\ref{4k1}) is indeed of the form in eq.~(\ref{gl21}).
All patterns for the neutrino spectrum are in principle possible. For a hierarchical spectrum $m_3>>m_2>>m_1$, $m_3^2 \sim \Delta m^2_{atm}$, $m_2^2/m_3^2 \sim \Delta m^2_{sol}/\Delta m^2_{atm}$ and $m_1$ could be negligible. But also degenerate masses and inverse hierarchy can be reproduced: for example, by taking $m_3= - m_2=m_1$  we have a degenerate model, while for $m_1= - m_2$ and $m_3=0$ an inverse hierarchy case is realized (stability under renormalization group running strongly prefers opposite signs for the first and the second eigenvalue which are related to solar oscillations and have the smallest mass squared splitting \cite{Antusch:2003kp,Antusch:2005gp,Mei:2005qp,Lindner:2005pk,Ellis:2005dr,Dighe:2007ksa,Lin:2009sq}). 

Note that the mass matrix for TB mixing, in the basis where charged leptons are diagonal, as given in eq.~(\ref{gl21}), can be specified as the most general matrix which is invariant under $\mu-\tau$ (or 2-3) symmetry and, in addition, under the action of a unitary symmetric matrix $S_{TB}$ (actually $S_{TB}^2=1$ and $[S_{TB},A_{23}]=0$):
\bea
m_\nu=S_{TB}m_\nu S_{TB}\,,\qquad\qquad
m_\nu=A_{23}m_\nu A_{23}\,,
\label{inv}
\eea
where $S_{TB}$ is given by
\bea
\label{trep}
S_{TB}&=\dd\frac{1}{3} \left(\begin{array}{ccc}
-1&2&2\\
2&-1&2\\
2&2&-1
\end{array}\right)\,.
\eea

Similarly, it is useful to consider the product $m^2=m_e^\dagger m_e$, where $m_e$ is the charged lepton mass matrix (defined as $\overline \psi_R m_e \psi_L$), because this product transforms as $m'^2=U_e^\dagger m^2 U_e$, with $U_e$ the unitary matrix that rotates the left-handed (LH) charged lepton fields. The most general diagonal $m^2$ is invariant under a diagonal phase matrix with 3 different phase factors,
\beq
m_e^\dagger\, m_e= T^\dagger \,m_e^\dagger\, m_e\, T\,,
\label{Tdiag}
\eeq
and conversely a matrix $m_e^\dagger m_e$ satisfying the above requirement is diagonal. If $T^n=1$
the matrix $T$ generates a cyclic group $Z_n$.
In the simplest case $n=3$ and we get $Z_3$ but $n>3$ is equally possible. In the $n=3$ case we have
\bea
\label{ta4}
T_{TB}=\left(\begin{array}{ccc}
1&0&0\\
0&\omega&0\\
0&0&\omega^2
\end{array}\right).
\eea
where $\omega^3=1$, so that $T_{TB}^3=1$.

We are now in a position to explain the role of finite groups and to formulate the general strategy to obtain the special mass matrix of TB mixing. We must find a group $G_f$ which, for simplicity, must be as small as possible but large enough to contain the $S$ and $T$ transformations. A limited number of products of $S$ and $T$ close a finite group $G_f$.  Hence the group $G_f$ contains the subgroups $G_S$ and $G_T$ generated by monomials in $S$ and $T$, respectively. We assume that the theory is invariant under the spontaneously broken symmetry described by $G_f$. Then we must arrange a breaking of $G_f$ such that, at leading order, $G_f$ is broken down to $G_S$ in the neutrino mass sector and down to $G_T$ in the charged lepton mass sector. In a good model this step must be realized in a natural way as a consequence of the stated basic principles, and not put in by hand. The symmetry under $A_{23}$ in some cases is also part of $G_f$ (this the case of $S_4$, the permutation group of 4 objects) and then must be preserved in the neutrino sector along with $S$ by the $G_f$ breaking or it could arise as a consequence of a special feature of the $G_f$ breaking (for example, in $A_4$ it is obtained by allowing only some transformation properties for the flavons with non vanishing VEV's). The explicit example of $A_4$ is discussed in the next section.
Note that, along the same line, a model with $\mu-\tau$ symmetry  can be realized in terms of the group $S_3$ generated by products of $A_{23}$ and $T$ (see, for example, Ref.~\cite{Feruglio:2007hi,Meloni:2012ci}).

%
%%%%%%%%%%%%%%%%%%%%%%%% 3.  The $A_4$ Group   %%%%%%%%%%%%%%%%
%	
\boldmath
\section{The $A_4$ Group}
\unboldmath
\label{sec:A4group}

$A_4$ is the group of the even permutations of 4 objects. It has 4!/2=12 elements. Geometrically, it can be seen as the invariance group of a tetrahedron (the odd permutations, for example the exchange of two vertices, cannot be obtained by moving a rigid solid). Let us denote a generic permutation $(1,2,3,4)\rightarrow (n_1,n_2,n_3,n_4)$ simply by $(n_1n_2n_3n_4)$. $A_4$ can be generated by two basic permutations $S$ and $T$ given by $S=(4321)$ and $T=(2314)$. One checks immediately that
\be\label{pres}
S^2=T^3=(ST)^3=1\,.
\ee
This is called a ``presentation'' of the group. The 12 even permutations belong to 4 equivalence classes ($h$ and $k$ belong to the same class if there is a $g$ in the group such that $ghg^{-1}=k$) and are generated from $S$ and $T$ as follows:
\beq
\begin{aligned}
&C_1: &&	I=(1234)\\ 
&C_2: &&	T=(2314),ST=(4132),TS=(3241),STS=(1423)\\ 
&C_3: &&	T^2=(3124),ST^2=(4213),T^2S=(2431),TST=(1342)\\ 
&C_4: &&	S=(4321),T^2ST=(3412),TST^2=(2143)
\label{class}
\end{aligned}
\eeq
Note that, except for the identity $I$ which always forms an equivalence class in itself, the other classes are according to the powers of $T$ (in $C_4$, $S$ could as well be seen as $ST^3$).

\begin{vchtable}[h!]
\vchcaption{Characters of $A_4$}
\label{tcarA4}
\begin{tabular}{@{}ccccc@{}}
\hline 
&&&&\\[-3mm]
\textbf{Class} 			& {\boldmath$\chi^1$} 	& {\boldmath$\chi^{1'}$} 	&{\boldmath$\chi^{1''}$}		&{\boldmath$\chi^3$}\\[1mm]
\hline
&&&&\\[-3mm] 
$C_1$ 			& 1 								& 1									& 1									&3\\[1mm]
$C_2$ 			& 1 								&$\omega$						&$\omega^2$					&0\\[1mm]
$C_3$ 			& 1 								& $\omega^2$					&$\omega$						&0\\[1mm]
$C_4$ 			& 1 								& 1									& 1									&-1\\[1mm]
\hline
\end{tabular}
\end{vchtable}

The characters of a group $\chi_g^R$ are defined, for each element $g$, as the trace of the matrix that maps the element in a given representation $R$. From the invariance of traces under similarity transformations it follows that equivalent representations have the same characters and that characters have the same value for all elements in an equivalence class. Characters satisfy $\sum_g \chi_g^R \chi_g^{S*}= N \delta^{RS}$, where $N$ is the number of transformations in the group ($N=12$ in $A_4$).  Also, for each element $h$, the character of $h$ in a direct product of representations is the product of the characters: $\chi_h^{R\otimes S}=\chi_h^R \chi_h^S$ and also is equal to the sum of the characters in each representation that appears in the decomposition of $R\otimes S$. In a finite group the squared dimensions of the inequivalent irreducible representations add up to $N$. The character table of $A_4$ is given in Tab.~\ref{tcarA4}. 
From this table one derives that $A_4$ has four inequivalent representations: three of dimension one, $1$, $1'$ and $1^{\prime\prime}$ and one of dimension $3$. 

It is immediate to see that the one-dimensional unitary representations are obtained by:
\beq
\begin{aligned}
&1	&\qquad&S=1	&\qquad&T=1\\ 
&1'	&\qquad&S=1	&\qquad&T=e^{\dd i 2 \pi/3}\equiv\omega\\
&1''	&\qquad&S=1	&\qquad&T=e^{\dd i 4\pi/3}\equiv\omega^2\,.
\end{aligned}
\label{uni}
\eeq
Note that $\omega=-1/2+i \sqrt{3}/2$ is the cubic root of 1 and satisfies $\omega^2=\omega^*$, $1+\omega+\omega^2=0$.

The three-dimensional unitary representation, in a basis
where the element $S=S'$ is diagonal, is built up from:
\beq
S'=\left(\begin{array}{ccc}
1&0&0\\
0&-1&0\\
0&0&-1
\end{array}\right)\,,\qquad\qquad
T'=\left(\begin{array}{ccc}
0&1&0\\
0&0&1\\
1&0&0
\end{array}\right)\,.
\label{tre}
\eeq

The multiplication rules are as follows: the product of two 3 gives $3 \times 3 = 1 + 1' + 1'' + 3 + 3$ and $1' \times 1' = 1''$, $1' \times 1'' = 1$, $1'' \times 1'' = 1'$ etc.
If $3\sim (a_1,a_2,a_3)$ is a triplet transforming by the matrices in eq.~(\ref{tre}) we have that under $S'$: $S'(a_1,a_2,a_3)^t= (a_1,-a_2,-a_3)^t$ (here the upper index $t$ indicates transposition)  and under $T'$: $T'(a_1,a_2,a_3)^t= (a_2,a_3,a_1)^t$. Then, from two such triplets $3_a\sim (a_1,a_2,a_3)$, $3_b\sim (b_1,b_2,b_3)$ the irreducible representations obtained from their product are:
\beq
\begin{gathered}
1=a_1b_1+a_2b_2+a_3b_3\\[1mm]
1'=a_1b_1+\omega^2 a_2b_2+\omega a_3b_3\\[1mm]
1^{\prime\prime}=a_1b_1+\omega a_2b_2+\omega^2 a_3b_3\\[1mm]
3\sim (a_2b_3, a_3b_1, a_1b_2)\\[1mm]
3\sim (a_3b_2, a_1b_3, a_2b_1)
\end{gathered}
\eeq
In fact, take for example the expression for $1^{\prime\prime}=a_1b_1+\omega a_2b_2+\omega^2 a_3b_3$. Under $S'$ it is invariant and under $T'$ it goes into $a_2b_2+\omega a_3b_3+\omega^2 a_1b_1=\omega^2[a_1b_1+\omega a_2b_2+\omega^2 a_3b_3]$ which is exactly the transformation corresponding to $1^{\prime\prime}$. 

In eq.~(\ref{tre}) we have the representation 3 in a basis where $S$ is diagonal. We shall see that for   our purposes it is convenient to go to a basis where instead it is $T$ that is diagonal. This is obtained through the unitary transformation:
\bea
T&=&VT'V^\dagger=\left(\begin{array}{ccc}
1&0&0\\
0&\omega&0\\
0&0&\omega^2
\end{array}\right)\,,\label{mT}\\
S&=&VS'V^\dagger=\frac{1}{3} \left(\begin{array}{ccc}
-1&2&2\\
2&-1&2\\
2&2&-1
\end{array}\right)\,.
\label{mS}
\eea
where:
\be \label {vu}
V=\frac{1}{\sqrt{3}} \left(\begin{array}{ccc}
1&1&1\\
1&\omega^2&\omega\\
1&\omega&\omega^2
\end{array}\right)\,.
\ee
The matrix $V$ is special in that it is a $3\times3$ unitary matrix with all entries of unit absolute value. It is interesting that this matrix was proposed long ago as a possible mixing matrix for neutrinos \cite{Cabibbo:1977nk, Wolfenstein:1978uw}. We shall see in the following that  in the $T$ diagonal basis the charged lepton mass matrix (to be precise the matrix $m_e^\dagger m_e$) is diagonal. Notice that the matrices $(S,T)$ of eqs.~(\ref{mT}) and (\ref{mS}) coincide
with the matrices $(S_{TB},T_{TB})$ of the previous section.

In this basis the product rules of two triplets, ($\psi_1,\psi_2,\psi_3$) and ($\varphi_1,\varphi_2,\varphi_3$) of $A_4$, according to the multiplication rule $3\times 3=1+1'+1^{\prime\prime}+3+3$ are different than in the $S$ diagonal basis (because for Majorana mass matrices the relevant scalar product is $(ab)$ and not  $(a^\dagger b)$)and are given by:
\beq
\begin{gathered}
\psi_1\varphi_1+\psi_2\varphi_3+\psi_3\varphi_2 \sim 1\,, \\
\psi_3\varphi_3+\psi_1\varphi_2+\psi_2\varphi_1 \sim 1'\,, \\
\psi_2\varphi_2+\psi_3\varphi_1+\psi_1\varphi_3 \sim 1''\,,\\
\left(\ba{c}
2\psi_1\varphi_1-\psi_2\varphi_3-\psi_3\varphi_2 \\
2\psi_3\varphi_3-\psi_1\varphi_2-\psi_2\varphi_1 \\
2\psi_2\varphi_2-\psi_1\varphi_3-\psi_3\varphi_1 \\
  \ea\right) \sim 3_S~, \qquad
\left(\ba{c}
\psi_2\varphi_3-\psi_3\varphi_2 \\
\psi_1\varphi_2-\psi_2\varphi_1 \\
\psi_3\varphi_1-\psi_1\varphi_3 \\
  \ea\right) \sim 3_A\,.
\end{gathered}
\label{tensorp}
\eeq

An obvious representation of $A_4$ is obtained by considering the $4\times4$ matrices that directly realize each permutation. For $S=(4321)$ and $T=(2314)$ we have
\be \label{trepp}
S_4= \left(\begin{array}{cccc}
0& 0& 0& 1\\
0& 0& 1& 0\\
0& 1& 0& 0\\
1& 0& 0& 0
\end{array}\right)\,,\qquad\qquad
T_4=\left(\begin{array}{cccc}
0& 1& 0& 0\\
0& 0& 1& 0\\
1& 0& 0& 0\\
0& 0& 0& 1
\end{array}\right)\,.
\ee

The matrices $S_4$ and $T_4$ satisfy the relations in eq.~(\ref{pres}), thus providing a representation
of $A_4$. Since the only irreducible representations of $A_4$ are a triplet and three singlets,
the $4\times4$ representation described by $S_4$ and $T_4$ is not irreducible. It decomposes
into the sum of the invariant singlet plus the triplet representation. In fact the vector $(1,1,1,1)^T$ is clearly invariant under permutations and similarly the 3-dimensional space orthogonal to it.
In matrix terms this decomposition is realized by the unitary matrix \cite{Altarelli:2006kg}
$U$ given by
\be \label {u43}
U=\frac{1}{2}
\left(\begin{array}{cccc}
+1&+1&+1&+1\\
-1&+1&+1&-1\\
+1&-1&+1&-1\\
+1&+1&-1&-1
\end{array}\right)\,.
\ee
This matrix maps $S_4$ and $T_4$ into matrices that 
are block-diagonal:
\be
U S_4 U^\dagger=
\left(
\begin{array}{c|ccc}
1& &0&\\
\hline
& & &\\
0& &S&\\
& & &
\end{array}
\right)\,,\qquad\qquad
U T_4 U^\dagger=
\left(
\begin{array}{c|ccc}
1& &0&\\
\hline
& & &\\
0& &T&\\
& & &
\end{array}
\right)\,,
\ee
where $S$ and $T$ are the generators of the three-dimensional representation in eq.~(\ref{tre}).
   
In the following we will work in the $T$ diagonal basis, unless otherwise stated. In this basis the 12 matrices of the 3-dimensional representation of $A_4$ are given as follows:
\begin{center}
\begin{tabular}{lll}
  $\cC_1$ : &$1= \left(
                 \begin{array}{ccc}
                   1 & 0 & 0 \\
                   0 & 1 & 0 \\
                   0 & 0 & 1 \\
                 \end{array}
               \right)$\,, \\
  \\
  $\cC_2$ : 
  				&$T=
               \left(
                 \begin{array}{ccc}
                   1 & 0 & 0 \\
                   0 & \om & 0 \\
                   0 & 0 & \om^2 \\
                 \end{array}
               \right)$\,,\qquad
                 &$ST=\dfrac{1}{3}\left(
                 \begin{array}{ccc}
                   -1 & 2\om & 2\om^2 \\
                   2 & -\om & 2\om^2 \\
                   2 & 2\om & -\om^2 \\
                 \end{array}
               \right)$\,,
\\[0.6cm]
             	&$TS=\dfrac{1}{3}\left(
                 \begin{array}{ccc}
                   -1 & 2 & 2 \\
                   2\om & -\om & 2\om \\
                   2\om^2 & 2\om^2 & -\om^2 \\
                 \end{array}
               \right)$\,, 
               	&$STS=\dfrac{1}{3}\left(
                 \begin{array}{ccc}
                   -1 & 2\om^2 & 2\om \\
                   2\om^2 & -\om & 2 \\
                   2\om & 2 & -\om^2 \\
                 \end{array}
               \right)$\,,
\\
\\
   $\cC_3$ : 
   				&$T^2=\left(
                 \begin{array}{ccc}
                   1 & 0 & 0 \\
                   0 & \om^2 & 0 \\
                   0 & 0 & \om \\
                 \end{array}
               \right)$\,,
              	&$ST^2=\dfrac{1}{3}\left(
                 \begin{array}{ccc}
                   -1 & 2\om^2 & 2\om \\
                   2 & -\om^2 & 2\om \\
                   2 & 2\om^2 & -\om \\
                 \end{array}
               \right)$\,,\\[0.6cm]
               &$T^2S=\dfrac{1}{3}\left(
                 \begin{array}{ccc}
                   -1 & 2 & 2 \\
                   2\om^2 & -\om^2 & 2\om^2 \\
                   2\om & 2\om & -\om \\
                 \end{array}
               \right)$\,,             
               &$TST=\dfrac{1}{3}\left(
                 \begin{array}{ccc}
                   -1 & 2\om & 2\om^2 \\
                   2\om & -\om^2 & 2 \\
                   2\om^2 & 2 & -\om \\
                 \end{array}
               \right)$\,,
\\
\\
$\cC_4$ : 
				&$S=\dfrac{1}{3}\left(
                 \begin{array}{ccc}
                   -1 & 2 & 2 \\
                   2 & -1 & 2 \\
                   2 & 2 & -1 \\
                 \end{array}
               \right)$\,,
               	&$T^2ST=\dfrac{1}{3}\left(
                 \begin{array}{ccc}
                   -1 & 2\om & 2\om^2 \\
                   2\om^2 & -1 & 2\om \\
                   2\om & 2\om^2 & -1 \\
                 \end{array}
               \right)$\,,\\[0.6cm]
               &$TST^2=\dfrac{1}{3}\left(
                 \begin{array}{ccc}
                   -1 & 2\om^2 & 2\om \\
                   2\om & -1 & 2\om^2 \\
                   2\om^2 & 2\om & -1 \\
                 \end{array}
               \right)$\,. 
\end{tabular}
\end{center}

We can now see why $A_4$ works for TB mixing. In Sec.~\ref{sec:NuPatters} we have already mentioned that the most general mass matrix for TB mixing in eq.~(\ref{gl21}), in the basis where charged leptons are diagonal, can be specified as one which is invariant under the 2-3 (or $\mu-\tau$) symmetry  and under the $S$ unitary transformation, as stated in eq.~(\ref{inv}). 
This observation plays a key role in leading to $A_4$ as a candidate group for TB mixing, because $S$ is a matrix of $A_4$. Instead the matrix $A_{23}$ is not an element of $A_4$ 
(because the 2-3 exchange is an odd permutation). 
We shall see that in $A_4$ models the 2-3 symmetry of the neutrino mass matrix arises as an accidental symmetry of the LO Lagrangian by imposing that there are no flavons transforming as $1'$ or $1''$ that break $A_4$ with two different VEV's (in particular one can assume that there are no flavons in the model transforming as $1'$ or $1''$).
It is also clear that a generic diagonal charged lepton matrix $m_e^\dagger m_e$ is characterized by the invariance under $T$, or $T^\dagger m_e^\dagger m_e T=m_e^\dagger m_e$.
	
The group $A_4$ has two obvious subgroups: $G_S$, which is a reflection subgroup
generated by $S$, and $G_T$, which is the group generated by $T$, which is isomorphic to $Z_3$.
If the flavour symmetry associated to $A_4$ is broken by the VEV of a triplet
$\varphi=(\varphi_1,\varphi_2,\varphi_3)$ of scalar fields,
there are two interesting breaking pattern. The VEV
\be
\langle\varphi\rangle=(v_S,v_S,v_S)
\label{unotre}
\ee
breaks $A_4$ down to $G_S$, while
\be
\langle\varphi\rangle=(v_T,0,0)
\label{unozero}
\ee
breaks $A_4$ down to $G_T$. As we will see, $G_S$ and $G_T$ are the relevant low-energy symmetries
of the neutrino and the charged-lepton sectors, respectively. Indeed we have already seen that the TB mass matrix is invariant under $G_S$
and a diagonal charged lepton mass $m_e^\dagger m_e$ is invariant under $G_T$.

%
%%%%%%%%%%%%%%%%%%%%%%%% 4.  Applying $A_4$ to lepton masses and mixings  %%%%%%%%%%%%%%%
%	
\boldmath
\section{Applying $A_4$ to Lepton Masses and Mixings}
\label{sec:A4Models}
\unboldmath

In the lepton sector a typical $A_4$ model works as follows \cite{Altarelli:2005yx}.  One assigns
leptons to the four inequivalent
representations of $A_4$: LH lepton doublets $l$ transform
as a triplet $3$, while the RH charged leptons $e^c$,
$\mu^c$ and $\tau^c$ transform as $1$, $1''$ and $1'$, respectively. 
These models can be realized both with and without a see-saw mechanism. In the first case there are three right-handed neutrinos transforming as a triplet of $A_4$, while in the second case the source of neutrino masses
is a set of higher dimensional operators violating the total lepton number. Here we consider a  see-saw realization, so we also introduce conjugate neutrino fields $\nu^c$ transforming as a triplet of $A_4$. The fact that LH lepton doublets $l$ and, in the see-saw case, also the RH neutrinos $\nu^c$, transform as triplets is crucial to realize the fixed ratios of mass matrix elements needed to obtain TB mixing. A drawback is that for the ratio $r$, defined by $\Delta m^2_{sun}/\Delta m^2_{atm}$, one would expect $\sqrt{r} \approx \mathcal{O}(1)$ to be compared with the experimental value is $\sqrt{r} \approx 0.2$, which implies a moderate fine-tuning. 

One adopts a supersymmetric (SUSY) context also to make contact 
with Grand Unification (flavour symmetries are supposed to act near the GUT scale\footnote{When the flavour symmetry is broken contextually with the electroweak one, such as in Refs.~\cite{Ma:2001dn,Lavoura:2007dw,Morisi:2009sc}, strong constraints from FCNC transitions are usually present \cite{Toorop:2010ex,Toorop:2010kt}, that can eventually rule out the model.}). In fact, as well known, SUSY is important in GUT's for offering a solution to the hierarchy problem, for improving coupling unification and for making the theory compatible with bounds on proton decay. But, in models of lepton mixing, SUSY also helps for obtaining the vacuum alignment, because the SUSY constraints are very strong and limit the form of the superpotential very much. Thus SUSY is not necessary but it is a plausible and useful ingredient. The flavour symmetry is broken by two sets of flavons $\Phi_e$ and $\Phi_\nu$, invariant under the SM gauge symmetry, that at the LO break $A_4$ down to $G_T$ and $G_S$, respectively. At this order $\Phi_e$ couples only to the charge lepton sector and $\Phi_\nu$ to the neutrino sector. Typically $\Phi_e$ and $\Phi_\nu$ include triplets and invariant singlets under $A_4$, but models with flavons transforming as $1'$ and $1''$ have also been considered \cite{Cooper:2011rh,Cooper:2012wf}.  For example $\Phi_e$ can consist of the triplet $\varphi_T$ with the vacuum alignment in eq.~(\ref{unozero}) and $\Phi_\nu$ can include the triplet $\varphi_S$ with the vacuum alignment in eq.~(\ref{unotre}) and two invariant singlets $\xi$, $\tilde{\xi}$.
Two Higgs doublets $h_{u,d}$, invariant under $A_4$, are also introduced. One can obtain  the observed hierarchy among $m_e$, $m_\mu$ and
$m_\tau$ by introducing an additional U(1)$_{FN}$ flavour symmetry \cite{Froggatt:1978nt} under
which only the  RH  lepton sector is charged (recently some models were proposed with a different VEV alignment such that the charged lepton hierarchies are obtained without introducing a $U(1)$ symmetry \cite{Lin:2008aj,Altarelli:2009kr}).
We recall that $U(1)_{FN}$ is a simple flavour symmetry where particles in different generations are assigned (in general) different values of an Abelian charge.  Also Higgs fields may get a non zero charge. When the symmetry is spontaneously broken the entries of mass matrices are suppressed if there is a charge mismatch and more so if the corresponding mismatch is larger.
We assign FN-charges $0$, $q$ and $2q$ to $\tau^c$, $\mu^c$ and
$e^c$, respectively. There is some freedom in the choice of $q$.
Here we take $q=2$.
By assuming that a flavon $\theta$, carrying
a negative unit of FN charge, acquires a VEV 
$\langle \theta \rangle/\Lambda\equiv\lambda_C<1$, where $\lambda_C\equiv\sin\theta_C$, the Yukawa couplings
become field dependent quantities $y_{e,\mu,\tau}=y_{e,\mu,\tau}(\theta)$
and we have
\be
y_\tau\approx \mathcal{O}(1)\,,\qquad\qquad
y_\mu\approx O(\lambda_C^2)\,,\qquad\qquad
y_e\approx O(\lambda_C^4)\,.
\ee
Had we chosen $q=1$, we would have needed  $\langle \theta \rangle/\Lambda$ of order $\lambda_C^2$, to reproduce the above result.
The superpotential term for lepton masses, $w_l$ is given by
\be
w_l=y_e e^c (\varphi_T l)+y_\mu \mu^c (\varphi_T l)'+
y_\tau \tau^c (\varphi_T l)''+ y (\nu^c l)+
(x_A\xi+\tilde{x}_A\tilde{\xi}) (\nu^c\nu^c)+x_B (\varphi_S \nu^c\nu^c)+\ldots
\label{wlss}
\ee
with dots denoting higher   
dimensional operators that lead to corrections to the LO   
approximation. In our notation, the product of 2 triplets  $(3 3)$ transforms as $1$, 
$(3 3)'$ transforms as $1'$ and $(3 3)''$ transforms as $1''$. 
To keep our formulae compact, we omit to write the Higgs and flavon  fields
$h_{u,d}$, $\theta$ and the cut-off scale $\Lambda$. For instance 
$y_e e^c (\varphi_T l)$ stands for $y_e e^c (\varphi_T l) h_d \theta^4/\Lambda^5$. The parameters of the superpotential $w_l$ are complex, in particular those responsible for the
heavy neutrino Majorana masses, $x_{A,B}$. Some terms allowed by the $A_4$ symmetry, such as the terms 
obtained by the exchange $\varphi_T\leftrightarrow \varphi_S$, 
(or the term $(\nu^c\nu^c)$) are missing in $w_l$. 
Their absence is crucial and, in each version of $A_4$ models, is
motivated by additional symmetries.

The LO superpotential in eq. (\ref{wlss}) leads to a diagonal mass matrix $m_e^{(0)}$ for the charged leptons\footnote{We absorbed in $y_f$ $(f=e,\mu,\tau)$ the appropriate factor of $\langle\theta\rangle/\Lambda$.}
\be
m_e^{(0)}=v_d
\left(
\begin{array}{ccc}
y_e&0&0\\
0&y_\mu&0\\
0&0&y_\tau
\end{array}
\right) \eta\qquad
\text{with}\qquad
\eta\equiv\frac{v_T}{\Lambda}\,,
\ee
and to a neutrino mass matrix $m_\nu^{(0)}$ of the same form as that of eq. (\ref{gl21}). As for the neutrino spectrum both normal and inverted hierarchies 
can be realized. It is interesting that $A_4$ models with the see-saw mechanism typically lead to a light neutrino spectrum which satisfies the sum rule (among complex masses):
\be
\frac{1}{m_3}=\frac{1}{m_1}-\frac{2}{m_2}\,.\\
\label{sumr}
\ee
A detailed discussion of a spectrum of this type can be found in Refs. \cite{Altarelli:2005yx,Bazzocchi:2009da,Altarelli:2009kr,Dorame:2011eb}.
The above sum rule gives rise to bounds on the lightest neutrino mass.
As a consequence, for example, the possible values of $|m_{ee}|$ are restricted. For normal hierarchy we have
\be
|m_{ee}|\approx \dd\frac{4}{3\sqrt{3}} \Delta m^2_{sun} \approx 0.007~{\rm eV}\,.
\label{meeno}
\ee
while for inverted hierarchy
\be
|m_{ee}|\ge  \dd\sqrt{\frac{\Delta m^2_{atm}}{8}} \approx 0.017~{\rm eV}\,.
\label{meeio}
\ee
In a completely general framework, without the restrictions imposed by the flavour symmetry,
$|m_{ee}|$ could vanish in the case of normal hierarchy. In this model $|m_{ee}|$ is always
different from zero, though its value for normal hierarchy is probably too small to be detected
in the next generation of $0\nu\beta\beta$ experiments.

In the leading approximation $A_4$ models lead to exact TB mixing.  In these models TB mixing is implied by the symmetry at the leading order approximation which is corrected by non-leading effects. Given the set of flavour symmetries and having specified the field content, the non leading corrections to TB mixing, arising from higher dimensional effective operators, can be evaluated in a well defined expansion. 

The departure from the LO approximation depends on the subleading contributions $\delta m_e^{(1)}$, $\delta m_\nu^{(1)}$, to the charged lepton and the neutrino mass matrices, respectively:
\be
m_e=m_e^{(0)}+\delta m_e^{(1)}+\dots\,,\qquad\qquad
m_{\nu}=m_{\nu}^{(0)}+\delta m_{\nu}^{(1)}+\dots\,, \\
\label{deltas}
\ee
which can vary according to the model considered.
In all models considered here \cite{Altarelli:2005yp,Altarelli:2005yx,Lin:2009bw,Altarelli:2009kr} the NLO corrections to the charged lepton mass matrix are of the following type:
\be
\delta m_e^{(1)}=
v_d
\left(
\begin{array}{ccc}
\cO(y_e)		&\cO(y_e)		&\cO(y_e)\nn\\
\cO(y_\mu)	&\cO(y_\mu)	&\cO(y_\mu)\nn\\
\cO(y_\tau)	&\cO(y_\tau)	&\cO(y_\tau)
\end{array}
\right)\,\eta\,\xi\,,
\ee
where $\xi$ is small adimensional parameter given by the ratio between the flavon VEVs and $\Lambda$. The transformation needed to diagonalize $m_e$ is $V_e^T m_e U_e=m_e^{diag}$ where
\be
U_e=
\left(
\begin{array}{ccc}
1&c^e_{12}\,\xi&c^e_{13}\,\xi\\
-c^{e*}_{12}\,\xi&1&c^e_{23}\,\xi\\
-c^{e*}_{13}\,\xi&-c^{e*}_{23}\,\xi&1
\end{array}
\right)\,.
\label{ue}
\ee
To discuss the NLO contribution to $m_\nu$ we distinguish two cases.

%%%%%%%%%%%%%%%%%%%%%%%%%   2.1.1  Typical $A_4$ models  
\boldmath
\subsection{Typical $A_4$ Models}
\unboldmath

\begin{figure}[h!]
 \centering
  \subfigure[Correlation between $\sin^2\theta_{12}$ and $\sin^2\theta_{13}$.]
   {\includegraphics[width=7.3cm]{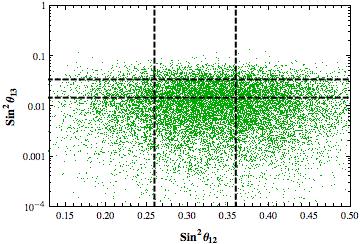}}
  \subfigure[Correlation between $\sin^2\theta_{23}$ and $\sin^2\theta_{13}$.]
   {\includegraphics[width=7.3cm]{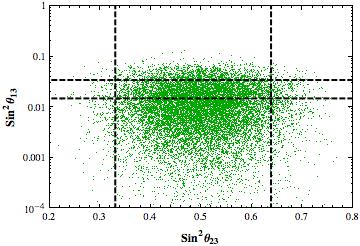}}
 \caption{ {\bf Typical {\boldmath$A_4$\unboldmath} Models.} On the left (right), we plot $\sin^2\theta_{13}$ as a function of $\sin^2\theta_{12}$ ($\sin^2\theta_{23}$), following eq.~(\ref{sinNLOTB}).  The dashed-black lines represent the $3\sigma$ values for the mixing angles from the Fogli {\it et al.} fit \cite{Fogli:2012ua}. Only the NH data sets is shown. The parameter $\xi$ is taken equal to $0.075$. The $c^{e,\nu}_{ij}$ parameters that multiply $\xi$ are treated as random complex numbers with absolute values following a Gaussian distribution around 1 with variance 0.5.}
 \label{fig:Sin12qvsSin13qTB_NIH}
\end{figure}

In ``typical'' $A_4$ models \cite{Altarelli:2005yx,Altarelli:2005yp,Altarelli:2009kr}, the NLO contribution $\delta m_\nu^{(1)}$ in eq.~(\ref{deltas}) is a generic symmetric matrix with entries suppressed, compared to the corresponding entries in $m_\nu^{(0)}$, by a relative factor $\xi'$,
of the order of the ratio between a flavon VEV and $\Lambda$. This occurs both with and without the see-saw mechanism. The generic transformation that diagonalizes $m_\nu$ is $U_\nu^T U_{TB}^T m_\nu U_{TB} U_\nu$ where
\beq
U_\nu=
\left(
\begin{array}{ccc}
1								&c^\nu_{12}\,\xi'				&c^\nu_{13}\,\xi'\\
-c_{12}^{\nu*}\,\xi'		&1									&c^\nu_{23}\,\xi'\\
-c_{13}^{\nu*}\,\xi'		&-c_{23}^{\nu*}\,\xi'			&1\\
\end{array}
\right)\,,
\label{unu}
\eeq
where $c^\nu_{12}$, $c^\nu_{13}$ and $c^\nu_{23}$ are complex parameters of order one in absolute value. Barring a fine-tuning of the Lagrangian parameters, in these models the suppression factors
$\xi$ and $\xi'$ are expected to be of the same order of magnitude. For example, beyond the LO the equations satisfied by $\langle\Phi_e\rangle$ and $\langle\Phi_\nu\rangle$ are no longer decoupled and the corrections
to the LO flavon VEVs turn out to be of the same size, for both $\Phi_e$ and $\Phi_\nu$. All the elements of the mixing matrix get corrections of the same size $\xi\approx \xi'$. We expect\footnote{Eq.~(\ref{sinNLOTB}) is a particular case of the general parametrization presented Ref.~\cite{King:2007pr}:
\beq
\sin\theta_{23}=\dfrac{1}{\sqrt2}(1+a)\,,\qquad
\sin\theta_{12}=\dfrac{1}{\sqrt3}(1+s)\,,\qquad
\sin\theta_{13}=\dfrac{r}{\sqrt2}\,,
\eeq
with $a$, $s$ and $r$ real numbers. The expressions in Eq.~(\ref{sinNLOTB}) show explicitly the dependence of the NLO mixing angles on the corrections from both the neutrino and the charged lepton sectors.} On the contrary, the expressions in eq.~(\ref{sinNLOTB}) show explicitly the dependence of the NLO mixing angles on the corrections from both the neutrino and charged lepton sectors.:
\beq
\begin{aligned}
\sin^2\theta_{23}&=\frac{1}{2}+{\cal R}e(c^e_{23})\,\xi+\dfrac{1}{\sqrt{3}}\left({\cal R}e(c^\nu_{13})-\sqrt2\,{\cal R}e(c^\nu_{23})\right)\,\xi\\
\sin^2\theta_{12}&=\frac{1}{3}-\frac{2}{3}{\cal R}e(c^e_{12}+c^e_{13})\,\xi+\dfrac{2\sqrt2}{3}\,{\cal R}e(c^\nu_{12})\,\xi\\[1mm]
\sin\theta_{13}&=\dfrac{1}{6}\left|3\sqrt2\left(c^e_{12}-c^e_{13}\right)+2\sqrt3\left(\sqrt2\,c^\nu_{13}+c^\nu_{23}\right)\right|\,\xi\,.
\end{aligned}
\label{sinNLOTB}
\eeq
According to these expressions, in order to reach the central value for the reactor angle in agreement with eq.~(\ref{ourTheta13}), the parameter $\xi$ is expected to be $\cO(0.1)$. A precise value can be found by studying the success rate to reproduce all the three mixing angles inside the corresponding $3\sigma$ ranges, depending on the value of $\xi$. As shown in ref. \cite{Altarelli:2012bn}, in a scan with the $c^{e,\nu}_{ij}$ parameters that multiply $\xi$ treated as random complex numbers with absolute values following a Gaussian distribution around 1 with variance 0.5, the value of $\xi$ that maximizes the success rate is found to be $0.075 (0.078)$ for the NH (IH). The corresponding success rate is $\sim 12 \%$, which is not large but not hopelessly small either. For this value of $\xi$ in Fig.~\ref{fig:Sin12qvsSin13qTB_NIH} we quantitatively analyze  the expressions in eq.~(\ref{sinNLOTB}) and their correlations:  in the plots on the left (right), we show the correlation between $\sin^2\theta_{13}$ and $\sin^2\theta_{12}$ ($\sin^2\theta_{23}$). In the plots we show only the NH case. The IH case is similar.

As we can see, the plots are representing the general behaviour of this class of models: $\sin^2\theta_{13}$ increases with $\xi$, but correspondingly also the deviation of $\sin^2\theta_{12}$ from $1/3$ does. As a result, even for the value of $\xi$ that maximizes the success rate, the requirement for having a reactor angle inside its $3\sigma$ error range corresponds to a prediction for the solar angle that spans all the $3\sigma$ experimental error bar and is often not even in agreement with the data.

%%%%%%%%%%%%%%%%%%%%%%%%%   2.1.2  Special $A_4$ models  
\boldmath
\subsection{Special $A_4$ Models}
\label{sec:SpecialA4}
\unboldmath

In these models $\delta m_\nu^{(1)}$  in eq.~(\ref{deltas}) is protected by the assumed symmetries so that it remains invariant under $S$ and its relative size, compared to $m_\nu^{(0)}$, can be bigger than $\xi$. For instance, in Ref. \cite{Lin:2009bw}
$\langle\Phi_e\rangle$ and $\langle\Phi_\nu\rangle$ satisfy decoupled equations up to NLO so that it is possible to achieve $\langle\Phi_e\rangle<\langle\Phi_\nu\rangle$. Moreover $\langle\Phi_\nu\rangle$ couples to charged leptons only at the NNLO so that the dominant source of 
corrections to the neutrino mixing pattern is $\delta m_\nu^{(1)}$.
Including these NLO corrections we have
\be
m_\nu=
\left(
\begin{array}{ccc}
x		&y-w			&y+w\\
y-w	&x+z+w	&y-z\\
y+w	&y-z			&x+z-w
\end{array}
\right)\,,
\label{generalmnu}
\ee
where $w$ represents the part of the NLO corrections that cannot be absorbed by a redefinition of $x$, $y$ and $z$. The parameter $w$ is smaller that $x$, $y$, $z$, but
not necessarily much smaller. The crucial property of $m_\nu$ \cite{Altarelli:2012bn} is that it is still invariant under the action of $S$ (but not any more of $A_{23}$):
\be
S^T m_\nu S=m_\nu\,.
\ee
Actually it can be directly proven that the matrix in eq.~(\ref{generalmnu}) is the most general one invariant under $S$. 
The matrix $m_\nu$ can be diagonalized in two steps. First we transform $m_\nu$ by a Tri-Bimaximal rotation:
\be
m_\nu'=U_{TB}^T m_\nu U_{TB}=
\left(
\begin{array}{ccc}
x-y&0 &\sqrt{3} w \\
0&x+2y &0 \\
\sqrt{3} w&0 &x-y+2z
\end{array}
\right)\,,
\ee
Second, we perform a unitary transformation in the (1,3) plane:
\be
V=\left(
\begin{array}{ccc}
\alpha 		& 0		&\xi'\\
0				&1		&0\\
-\xi'^*		&0		&\alpha^*
\end{array}
\right)\,,\qquad\qquad\qquad|\alpha|^2+|\xi'|^2=1\,,
\label{exact13}
\ee
\be
V^T m_\nu' V= m_\nu^{diag}
\ee
The exact rotation is given by:
\be
\frac{2\alpha\xi'}{|\alpha|^2-|\xi'|^2}=\frac{u v^*(u^*-v)}{|v|^2-|u|^2}\,,\qquad
\text{with}\quad
u\equiv \frac{2\sqrt{3} w}{x-y}\,,\quad
\text{and}\quad
v\equiv-\frac{2 \sqrt{3} w}{x-y+2z}\,,
\ee
The unitary matrix that diagonalizes $m_\nu$ is
\be
U_{TB} V=
\left(
\begin{array}{ccc}
\sqrt{2/3}\alpha&1/\sqrt{3} &\sqrt{2/3} \xi' \\
-\alpha/\sqrt{6}+\xi'^*/\sqrt{2}&1/\sqrt{3} &-\alpha^*/\sqrt{2}-\xi'/\sqrt{6} \\
-\alpha/\sqrt{6}-\xi'^*/\sqrt{2}&1/\sqrt{3} &+\alpha^*/\sqrt{2}-\xi'/\sqrt{6}
\end{array}
\right)\,.
\label{specialmixing}
\ee
It is not restrictive to choose $\alpha$ real and positive and we have:
In eq. (\ref{specialmixing}) it is not restrictive to choose $\alpha$ real and positive and we have:
\bea
\delta_{CP}&\approx&\arg\xi'\\
\sin\theta_{13}&=&\left|\sqrt{\frac{2}{3}}\xi'+\frac{c^e_{12}-c^e_{13}}{\sqrt{2}}\xi\right|\\
\sin^2\theta_{12}&=&\frac{1}{3-2 |\xi'|^2}-\frac{2}{3}{\cal R}e(c^e_{12}+c^e_{13})\,\xi\,=\frac{1}{3}+\frac{2}{9}|\xi'|^2-\frac{2}{3}{\cal R}e(c^e_{12}+c^e_{13})\,\xi
\label{sinNNLOTBLinSol}\\
\sin^2\theta_{23}&=&\frac{1}{2}
\frac{\left(1+\frac{\xi'}{\sqrt{3}\alpha}\right)\left(1+\frac{\xi'^*}{\sqrt{3}\alpha}\right)}
{\left(1+\frac{|\xi'|^2}{3\alpha^2}\right)}+{\cal R}e(c^e_{23})\,\xi\,=\frac{1}{2}+\frac{1}{\sqrt{3}}|\xi'|\cos\delta+{\cal R}e(c^e_{23})\,\xi
\label{sinNNLOTBLinAtm}
\eea
where we have also included the effects coming from the diagonalization of the charged lepton sector, to first order in $\xi$. The second equality shows the result expanded in powers of $|\xi'|$, to the order $|\xi'|^2$. In these models $|\xi'|$ is of order $0.1$, bigger than $\xi$ so that the contribution of eq.~(\ref{ue}) are subdominant.

It is interesting to note that if we neglect the corrections proportional to $\xi$, we have an exact relation between the solar and the reactor angle\footnote{It has been shown in Ref.~\cite{Hernandez:2012ra}, from general group theoretical considerations, that these correlations are a general feature of flavour models when the symmetry group of the charged lepton (neutrino) mass matrix is $Z_3$ ($Z_2$).}:
\be
\sin^2\theta_{12}=\frac{1}{3(1-\sin^2\theta_{13})}\,,\qquad\qquad
\sin^2\theta_{23}=\frac{1}{2}+\frac{1}{\sqrt2}\sin\theta_{13}\cos\delta_{CP}\,.
\ee
The first expression demonstrates that the unitary transformation $V$ always increases the solar angle from the TB value, while the preferred 1$\sigma$ interval is below the TB prediction. This is a small effect, of second order in $\theta_{13}$, that can be compensated by the corrections proportional to $\xi$. The second correlation involves the Dirac CP phase and is particularly interesting considering the recent hint of a CP phase close to $\pi$ for the NH case: when considering the $1\sigma$ ($2\sigma$) ranges for the mixing angles, one sees an indication that $\cos\delta_{CP}$ lies in the interval $[-1,-0.5]$, while no indication arises when the $3\sigma$ error band for $\sin^2\theta_{23}$ is taken into account. Although these results for the CP phase is modified by the inclusion of the subleading $\xi$ contributions, these correlations will allow an interesting test for such models once $\delta_{CP}$ is measured and the precision on $\sin^2\theta_{23}$ is improved.

\begin{figure}[h!]
 \centering
   \subfigure[Correlation between $\sin^2\theta_{12}$ and $\sin^2\theta_{13}$.]
   {\includegraphics[width=7cm]{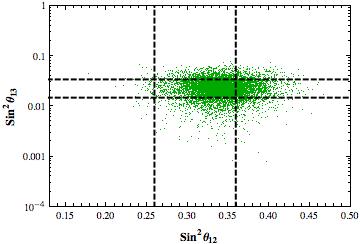}}
   \subfigure[Correlation between $\sin^2\theta_{23}$ and $\sin^2\theta_{13}$.]
   {\includegraphics[width=7cm]{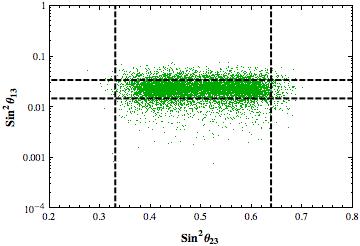}}
 \caption{ {\bf Special {\boldmath$A_4$\unboldmath} Models.} $\sin^2\theta_{13}$ as a function of $\sin^2\theta_{12}$ ($\sin^2\theta_{23}$ is plotted on the left (right), following eqs.~(\ref{sinNNLOTBLinSol}) and (\ref{sinNNLOTBLinAtm}).  The dashed-black lines represent the $3\sigma$ values for the mixing angles from the Fogli {\it et al.} fit \cite{Fogli:2012ua}. Only the NH data sets is shown. The parameter $\xi$ is a real number in $[0.005,\,0.06]$; $\xi'$ is a complex number with absolute values equal to $0.183$; the parameters $c^e_{ij}$ are random complex numbers with absolute values following a Gaussian distribution around 1 with variance 0.5.}
 \label{fig:Sin23qvsSin13qTBLin_NIH}
\end{figure}

The success rate to reproduce all the three mixing angles inside their corresponding $3\sigma$ error ranges, as a function of $|\xi'|$, is studied in ref. \cite{Altarelli:2012bn}. The parameters are chosen such that $\xi$ is a real number in $[0.005,\,0.06]$ and $c^e_{ij}$ are random complex numbers with absolute values following a Gaussian distribution around 1 with variance 0.5. The value of $|\xi'|$ that maximizes the success rate for both the hierarchies is found to be $0.183$. The corresponding success rate is much larger in these models ($\sim 64 \%$) than for the typical $A_4$ models. For the stated range of $\xi$ and the optimal value of $\xi'$
the deviations in eqs.~(\ref{sinNNLOTBLinSol}) and (\ref{sinNNLOTBLinAtm}) and their correlations are quantitatively  analyzed in Fig.~\ref{fig:Sin23qvsSin13qTBLin_NIH}: in the plots on the left (right) column, we show the correlations in eqs.~(\ref{sinNNLOTBLinSol}) and (\ref{sinNNLOTBLinAtm}) between $\sin^2\theta_{13}$ and $\sin^2\theta_{12}$ or $\sin^2\theta_{23}$, respectively. We see that, for this choice of the parameters, the model can well describe all three angles inside the corresponding $3\sigma$ interval, and its success rate is much larger than that of the typical TB models.

One particularly interesting example realizing this scenario is provided by the Lin model \cite{Lin:2009bw} (see also \cite{Varzielas:2010mp}), formulated before the T2K, MINOS, DOUBLE CHOOZ, Daya Bay and RENO results were known.  In the Lin model, the $A_4$ symmetry breaking is arranged, by suitable additional $Z_n$ parities, in such a way that, not only at LO but also at NLO, the corrections to the charged lepton and the neutrino sectors are kept separate. Then the contributions to neutrino mixing from the diagonalization of the charged leptons can be of $\mathcal{O}(\lambda_C^2)$ while those in the neutrino sector can be of $\mathcal{O}(\lambda_C)$. In addition, in the Lin model these large corrections do not affect $\theta_{12}$ and satisfy the relations in eq. (\ref{sinNNLOTBLinAtm}), with $\delta$ being the CKM-like CP violating phase of the lepton sector. Thus in the Lin model the NLO corrections to the solar angle $\theta_{12}$ and to the reactor angle $\theta_{13}$ can naturally be of different orders and  $\theta_{13}\sim\mathcal{O}(\lambda_C)$ is not at all surprising.

A related scenario is provided by a framework based on a typical $A_4$ model as described at the beginning of this section, extended by the inclusion in $\Phi_\nu$ of two additional singlets transforming as $1'$ and $1''$ \cite{King:2011zj}. Once these singlets and  the triplet $\varphi_S$ acquire a VEV, the resulting LO neutrino mass matrix $m_\nu^{(0)}$ is only invariant under the action of $S$
and no more under $A_{23}$. Thus, already at the LO we have $m_\nu^{(0)}$
of the general form in eq. (\ref{generalmnu}). In this framework the smallness of $\theta_{13}$ is however unexplained.

\boldmath
\subsection{Summary on $A_4$ models for lepton mixing}
\label{sec:SummaryA4}
\unboldmath

In summary, in the absence of specific dynamical tricks, in a generic $A_4$ model, all the three mixing angles receive corrections of the same order of magnitude. Since the experimentally allowed departures of $\theta_{12}$ from the TB value, $\sin^2{\theta_{12}}=1/3$, are small, numerically not larger than $\mathcal{O}(\lambda_C^2)$, it follows that both $\theta_{13}$ and the deviation of $\theta_{23}$ from the maximal value are expected to also be typically of the same general size. The central values $\sin\theta_{13} \approx 0.15$ that can be derived from the experimental results in Tab.~\ref{tab:t13} are in between  $\mathcal{O}(\lambda_C^2) \sim \mathcal{O}(0.05)$ and $\mathcal{O}(\lambda_C) \sim \mathcal{O}(0.23)$. Although models based on TB (or GR) mixing tend to lead to a smaller value of $\theta_{13}$ one can argue that they are still viable with preference for the lower side of the experimental range. But, as we have seen, one can introduce some additional theoretical input to enhance the value of $\theta_{13}$ in an $A_4$ model (several models have been recently proposed in order to fulfill this goal \cite{Xing:2011at,Zheng:2011uz,Ma:2011yi,Zhou:2011nu,Araki:2011wn,Haba:2011nv,Morisi:2011pm,Chao:2011sp,Zhang:2011aw,Dev:2011bd,Chu:2011jg,BhupalDev:2011gi,Toorop:2011jn,Antusch:2011qg,Rodejohann:2011uz,Ahn:2011if,King:2011zj,Marzocca:2011dh,Ge:2011qn,Kumar:2011vf,Bazzocchi:2011ax,Araki:2011qy,Antusch:2011ic,Fritzsch:2011qv,Rashed:2011zs,Ludl:2011vv,Verma:2011kz,Meloni:2011ac,Dev:2011hf,Deepthi:2011sk,Rashed:2011xe,deAdelhartToorop:2011re,deMedeirosVarzielas:2011wx,Araki:2011zg,Gupta:2011ct,Ding:2012xx,Ishimori:2012gv,Dev:2012ns,Bazzocchi:2012ve,BhupalDev:2012nm,Cooper:2012wf,Siyeon:2012zu,Wu:2012ri,Branco:2012vs,Meloni:2012ci,Ahn:2012tv,Varzielas:2012ss,Hernandez:2012ra,Hagedorn:2012pg,Hagedorn:2012ut}). As a result we now have examples of $A_4$ models where the departures from exact TB mixing are naturally larger for $\theta_{13}$ than for $\theta_{12}$ and $\theta_{23}$.

%
%%%%%%%%%%%%%%%%%%%%%%%% 5.  $A_4$, Quarks and GUT's  %%%%%%%%%%%%%%%
%	
\boldmath
\section{$A_4$, Quarks and GUT's}
\unboldmath
\label{sec:Extensions}

Much attention  has been devoted to the question whether models with TB mixing in the neutrino sector can be  suitably extended to also successfully describe the observed pattern of quark mixings and masses and whether this more complete framework can be made compatible with (supersymmetric) SU(5) or SO(10) Grand Unification. 

The simplest attempts of directly extending models based on $A_4$ to quarks have not been  satisfactory.
At first sight the most appealing
possibility is to adopt for quarks the same classification scheme under $A_4$ that one has
used for leptons (see, for example, Ref.\cite{Altarelli:2005yx}). Thus one tentatively assumes that LH quark doublets $Q$ transform
as a triplet $3$, while the  antiquarks $(u^c,d^c)$,
$(c^c,s^c)$ and $(t^c,b^c)$ transform as $1$, $1''$ and $1'$, respectively. This leads to $V_u=V_d$ and to the identity matrix for $V_{CKM}=V_u^\dagger V_d$ in the lowest approximation. This at first appears as very promising: a LO approximation where neutrino mixing is TB and $V_{CKM}=1$ is a very good starting point. But there are some problems. First, the corrections 
to $V_{CKM}=1$ turn out to be strongly constrained by the leptonic sector, because lepton mixing angles are very close to the TB values, and, in the simplest models, this constraint leads to a too small $V_{us}$
(i.e. the Cabibbo angle is rather large in comparison to the allowed shifts from the TB mixing angles). Also in these models, the quark classification which leads to $V_{CKM}=1$ is not compatible with $A_4$ commuting with SU(5). 
An additional consequence of the above assignment is that the top quark mass arises from a non-renormalizable dimension-5 operator. In that case, to reproduce the top mass, we need 
to compensate the cutoff suppression by some extra dynamical mechanism. Alternatively, we have to introduce a separate symmetry breaking parameter for the quark sector, sufficiently close to the cutoff
scale.

Due to this, larger discrete groups have been considered for the description of quarks.
A particularly appealing set of models is based on the discrete group $T'$, the double covering group of $A_4$ \cite{Carr:2007qw,Feruglio:2007uu,Chen:2007afa,Frampton:2007et,Aranda:2007dp,Ding:2008rj,Frampton:2009fw,Chen:2009gf,Merlo:2011hw}. The 
representations of $T'$ are those of $A_4$ plus three independent doublets 2, $2'$ and $2''$. The doublets are interesting for the classification of the first two generations of quarks \cite{Barbieri:1995uv,Barbieri:1996ww,Barbieri:1997tu}. For example, in Ref.\cite{Feruglio:2007uu} a viable description was obtained, i.e. in the leptonic sector the predictions of the $A_4$ model are maintained, while the $T'$ symmetry plays an essential role for reproducing the pattern of quark mixing. But, again, the classification adopted in this model is not compatible with Grand Unification.

As a result, the group $A_4$ was considered by many authors to be too
limited to also describe quarks and to lead to a grand unified
description. But it has been shown \cite{Altarelli:2008bg} that this negative attitude
is not justified and that it is actually possible to construct a
viable model based on $A_4$ which leads to a grand
unified theory (GUT) of quarks and leptons with TB mixing
for leptons and with quark (and charged lepton) masses and mixings compatible with experiment. At the same time this model offers an example of an
extra dimensional SU(5) GUT in which a description of all fermion masses
and mixings is accomplished.  The
formulation of SU(5) in extra dimensions has the usual advantages of
avoiding large Higgs representations to break SU(5) and of solving the
doublet-triplet splitting problem.  The choice of the transformation properties of the two
Higgses $H_5$ and $H_{\overline{5}}$ has a special role in this model. They are chosen to transform 
as two different $A_4$ singlets
$1$ and $1'$. As a consequence, mass terms for the Higgs colour
triplets are  not directly allowed and their masses are
introduced by orbifolding, \`{a} la Kawamura \cite{Kawamura:2000ev}.  In this model, proton
decay is dominated by gauge vector boson exchange giving rise to
dimension-6 operators, while the usual contribution of dimension-5 operators is forbidden by the selection rules of the model. Given the large $M_{GUT}$ scale of SUSY models and the relatively huge theoretical uncertainties, the decay rate is within the present experimental limits.
A see-saw realization
in terms of an $A_4$ triplet of RH neutrinos $\nu^c$ ensures the
correct ratio of light neutrino masses with respect to the GUT
scale. In this model extra dimensional effects directly
contribute to determine the flavour pattern, in that the two lightest
tenplets $T_1$ and $T_2$ are in the bulk (with a doubling $T_i$ and
$T'_i$, $i=1,2$ to ensure the correct zero mode spectrum), whereas the
pentaplets $F$ and $T_3$ are on the brane. The hierarchy of quark and
charged lepton masses and of quark mixings is determined by a
combination of extra dimensional suppression factors and of $U(1)_{FN}$ charges, both of which only apply to the first two
generations, while the neutrino mixing angles
derive from $A_4$ in the usual way. If the extra dimensional suppression factors and the $U(1)_{FN}$ charges are switched off, only the third generation masses of quarks and charged leptons survive. Thus the charged fermion mass matrices are nearly empty in this limit (not much of $A_4$ effects remain) and the quark mixing angles are determined by the small corrections induced by the above effects. The model is natural, since most of the
small parameters in the observed pattern of masses and mixings as well
as the necessary vacuum alignment are  justified by the symmetries of
the model. However, in this case, like in all models based on $U(1)_{FN}$, the number of $\mathcal{O}(1)$ parameters is larger than the number of measurable quantities, so that in the quark sector the model can only account for the orders of magnitude (measured in terms of powers of an expansion parameter) and not for the exact values of mass ratios and mixing angles. A moderate fine-tuning is only needed to enhance the Cabibbo mixing angle between the first two generations, which would generically be of $\mathcal{O}(\lambda_C^2)$. 

The problem of constructing GUT models based on  $SU(5)\otimes G_f$ or $SO(10)\otimes G_f$ with approximate TB mixing in the leptonic sector has also been considered by many authors. Examples are: for $G_f=A_4$ Ref.\cite{Ma:2005tr,Ma:2006sk,Ma:2006wm,Morisi:2007ft,Grimus:2008tm,Altarelli:2008bg,Ciafaloni:2009ub,Bazzocchi:2008rz,Antusch:2010es}, for $T'$ Ref.\cite{Chen:2007afa,Chen:2009gf}, for $S_4$ Ref.\cite{Ishimori:2010xk,Hagedorn:2010th,Ding:2010pc,Antusch:2011sx}. 
As for the models based on $SO(10)\otimes G_f$  recent examples were discussed with $G_f=S_4$ \cite{Dutta:2009ij,Adulpravitchai:2010na,Patel:2010hr,BhupalDev:2012nm} and $G_f=PSL_2(7)$ \cite{King:2009mk,King:2009tj}. Clearly the case of $SO(10)$ is even more difficult than that of $SU(5)$ because the neutrino sector is tightly related to that of quarks and charged leptons as all belong to the 16 of $SO(10)$. For a discussion of $SO(10)\otimes A_4$ models, see \cite{Bazzocchi:2008sp}. More in general see Refs.\cite{Altarelli:2010at,Joshipura:2011rr,Joshipura:2011nn,BhupalDev:2011gi,Blankenburg:2011vw}.
In our opinion most of the models are incomplete (for example, the crucial issue of VEV alignment is not really treated in depth as it should) and/or involve a number of unjustified steps and ad-hoc fine-tuning of parameters. 

%
%%%%%%%%%%%%%%%%%%%%%%%% 6.  Possible Origin of $A_4$  %%%%%%%%%%%%%%%
%
\boldmath
\section{Possible Origin of $A_4$}
\unboldmath
\label{sec:A4origin}

There is an interesting relation \cite{Altarelli:2005yx} between the $A_4$ model considered so far and the modular group. This relation could possibly be relevant to understand the origin of the $A_4$ symmetry from a more fundamental layer of the theory.
The modular group $\Gamma$ is the group of linear fractional transformations acting on a complex variable $z$:
\be
z\to\frac{az+b}{cz+d}\,,\qquad\qquad ad-bc=1\,,
\label{frac}
\ee
where $a,b,c,d$ are integers. 
There are infinite elements in $\Gamma$, but all of them can be generated by the two
transformations:
\be
s:\,\, z\to -\frac{1}{z}\,,\qquad\qquad t:\,\, z\to z+1\,,
\label{st}
\ee
The transformations $s$ and $t$ in (\ref{st}) satisfy the relations
\be
s^2=(st)^3=1
\label{absdef}
\ee
and, conversely, these relations provide an abstract characterization of the modular group.
Since the relations in eqs.~(\ref{pres}) are a particular case of the more general constraint in eq.~(\ref{absdef}),
it is clear that $A_4$ is a very small subgroup of the modular group and that the $A_4$ representations discussed above are also representations of the modular group.
In string theory the transformations in eq.~(\ref{st})
operate in many different contexts. For instance the role of the complex 
variable $z$ can be played by a field, whose VEV can be related to a physical
quantity like a compactification radius or a coupling constant. In that case
$s$ in eq.~(\ref{st}) represents a duality transformation and $t$ in eq.~(\ref{st}) represents the transformation associated to an ``axionic'' symmetry. 

A different way to  
understand the dynamical origin of $A_4$ was presented in Ref. \cite{Altarelli:2006kg} where it is shown that the $A_4$ symmetry can be simply  
obtained by orbifolding starting from a model in 6 dimensions (6D).  
In this approach $A_4$ appears as the remnant of the reduction
from 6D to 4D space-time symmetry induced by the 
special orbifolding adopted.  
There are 4D branes at the four fixed points of the orbifolding and the  
tetrahedral symmetry of $A_4$ connects these branes. The standard  
model fields have components on the fixed point branes while the scalar  
fields necessary for the $A_4$ breaking are in the bulk. Each brane field, either a 
triplet or a singlet, has components on all of the four fixed points (in particular all components are 
equal for a singlet) but the interactions are local, i.e. all vertices involve products of field 
components at the same space-time point. This approach suggests a deep relation between flavour symmetry 
in 4D and  space-time symmetry in extra dimensions.

\begin{figure}[h!]
\includegraphics[width=10.0 cm]{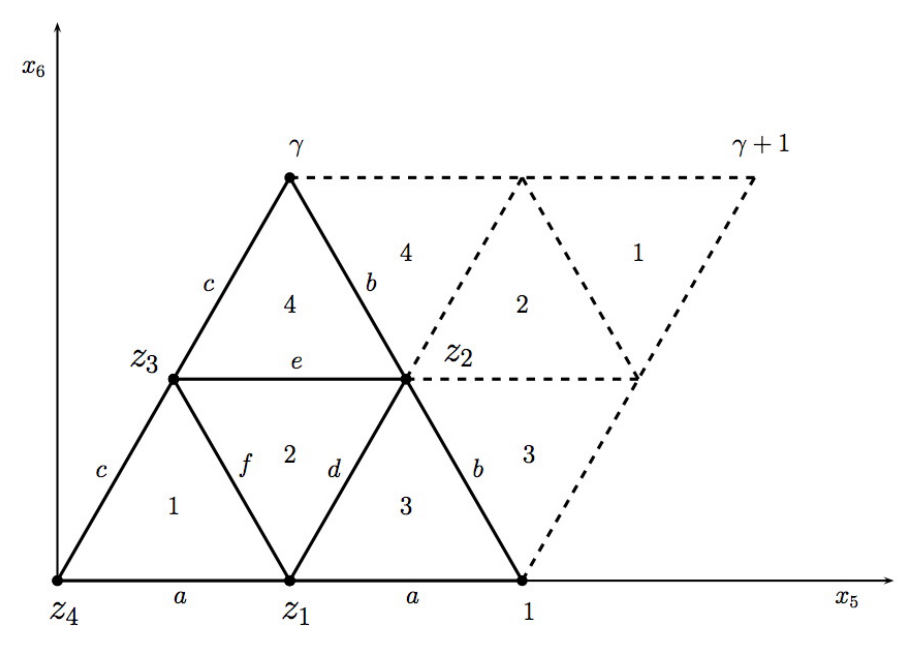}
\caption{Orbifold $T_2/Z_2$. The regions with the same numbers are 
identified with each other. The four triangles bounded by solid lines form the
fundamental region, where also the edges with the same letters are identified.
The orbifold $T_2/Z_2$ is exactly a regular tetrahedron with 6 edges
$a,b,c,d,e,f$ and four vertices $z_1$, $z_2$, $z_3$, $z_4$, corresponding to 
the four fixed points of the orbifold.}
\label{fig:tetrahedron}
\end{figure}

The orbifolding is defined as follows.
We consider a quantum field theory in 6 dimensions, with two extra dimensions
compactified on an orbifold $T^2/Z_2$. We denote by $z=x_5+i\, x_6$ the complex
coordinate describing the extra space. The torus $T^2$ is defined by identifying 
in the complex plane the points related by
\be
\begin{aligned}
&z\to z+1\\
&z\to z+\gamma\qquad\qquad
\gamma=e^{i\,\pi/3}\,,
\label{torus}
\end{aligned}
\ee
where our length unit, $2\pi R$, has been set to 1 for the time being.
The parity $Z_2$ is defined by
\be
z\to -z
\label{parity}
\ee
and the orbifold $T^2/Z_2$ can be represented by the fundamental region given by the triangle
with vertices $0,1,\gamma$, see Fig. \ref{fig:tetrahedron}. The orbifold has four fixed points, $(z_1,\,z_2,\,z_3,\,z_4)=\mbox{$(1/2,\,(1+\gamma)/2,\,\gamma/2,\,0)$}$.
The fixed point $z_4$ is also represented by the vertices $1$ and $\gamma$. In the orbifold,
the segments labelled by $a$ in Fig. \ref{fig:tetrahedron}, $(0,1/2)$ and $(1,1/2)$, are 
identified and similarly for those labelled by $b$, $(1,(1+\gamma)/2)$ and 
$(\gamma,(1+\gamma)/2)$, and those labelled by $c$, $(0,\gamma/2)$, $(\gamma,\gamma/2)$. Therefore the orbifold is a regular tetrahedron
with vertices at the four fixed points.

The symmetry of the uncompactified 6D space time is broken
by compactification. Here we assume that, before compactification,
the space-time symmetry coincides with the product of 6D translations
and 6D proper Lorentz transformations. The compactification breaks
part of this symmetry.
However, due to the special geometry of our orbifold, 
a discrete subgroup of rotations and translations in the extra space is left
unbroken. This group can be generated by two transformations:
\be
\begin{aligned}
{\cal S}:& z\to z+\frac{1}{2}\\
{\cal T}:& z\to \omega z \qquad\qquad\qquad
\omega\equiv\gamma^2~\,.
\end{aligned}
\label{rototra}
\ee
Indeed ${\cal S}$ and ${\cal T}$ induce even permutations of the four fixed points:
\be
\begin{aligned}
{\cal S}:& (z_1,z_2,z_3,z_4)\to (z_4,z_3,z_2,z_1)\\
{\cal T}:& (z_1,z_2,z_3,z_4)\to (z_2,z_3,z_1,z_4)\,,
\end{aligned}
\label{stfix}
\ee
thus generating the group $A_4$.  From 
the previous equations we immediately verify that ${\cal S}$ and ${\cal T}$ satisfy
the characteristic relations obeyed by the generators of $A_4$:
${\cal S}^2={\cal T}^3=({\cal ST})^3=1$.
These relations are actually satisfied not only at the fixed points, but on the whole orbifold,
as can be easily checked from the general definitions of ${\cal S}$ and ${\cal T}$ in eq.~(\ref{rototra}),
with the help of the orbifold defining rules in eqs.~(\ref{torus}) and (\ref{parity}).
We can exploit this particular geometry of the internal space to build a model 
with $A_4$ flavour symmetry.
There are 4D branes at the four fixed points of the orbifolding and the  
tetrahedral symmetry of $A_4$ connects these branes. The standard  
model fields have components on the fixed point branes while the scalar  
fields necessary for the $A_4$ breaking are in the bulk. Each brane field, either a 
triplet or a singlet, has components on all of the four fixed points (in particular all components are 
equal for a singlet) but the interactions are local, i.e. all vertices involve products of field 
components at the same space-time point. 
In the low-energy limit this model coincides with one of those presented in sec.~\ref{sec:A4Models} \cite{Altarelli:2005yx}.
Unfortunately in such a limit the 6D construction does not provide additional constraints or predictions.

This construction can be embedded in a SU(5) GUT \cite{Burrows:2009pi}.
Other discrete groups can arise from the compactification of two extra dimensions on orbifolds and the possibilities have been 
classified in Ref.~\cite{Adulpravitchai:2009id} within a field theory approach. In string theory the flavour symmetry can be larger than
the isometry of the compact space. For instance in heterotic orbifold models the orbifold geometry combines with the space group selection 
rules of the string, as shown in Ref.~\cite{Kobayashi:2006wq}. Discrete flavour symmetries from magnetized/intersecting D-branes are discussed in Ref.~\cite{Abe:2009vi}.
Discrete symmetries can also arise from the spontaneous breaking of continuous ones. Such a possibility has been discussed in Refs. \cite{Adulpravitchai:2009kd,Berger:2009tt,Luhn:2011ip}.

%
%%%%%%%%%%%%%%%%%%%%%%%%%%%   7.  Alternative routes to TB mixing  %%%%%%%%%%%%%%%%%%%
%
\section{Alternative routes to TB mixing}
\label{sec:Alternatives}

While $A_4$ is the minimal flavour group leading to TB mixing, alternative flavour groups have been studied in the literature and can lead to interesting variants with some specific features.

In Ref. \cite{Lam:2008sh}, the claim was made that, in order to obtain the TB mixing ``without fine-tuning'', the finite group must be $S_4$ or a larger group containing $S_4$. For us this claim is not well grounded being based on an abstract mathematical criterium for a natural model (see also Ref.~\cite{Grimus:2009pg}). For us a physical field theory model is natural if the interesting results are obtained from the most general lagrangian compatible with the stated symmetry and the specified representation content for the flavons. For example, in Ref.~\cite{Altarelli:2005yp,Altarelli:2005yx}, a natural (in our sense) model for the TB mixing is built with $A_4$ (which is a subgroup of $S_4$) by simply not including symmetry breaking flavons transforming like the $1'$ and the $1''$ representations of $A_4$. This limitation on the transformation properties of the flavons is not allowed by the rules specified in Ref. \cite{Lam:2008sh}, which demands that the symmetry breaking is induced by all possible kinds of flavons (note that, according to this criterium, the SM of electroweak interactions would not be natural because only Higgs doublets are introduced!). Rather, for naturalness we also require that additional physical properties like the VEV alignment or the hierarchy of charged lepton masses also follow from the assumed symmetry and are not obtained by fine-tuning parameters: for this actually $A_4$ can be more effective than $S_4$ because it possesses three different singlet representations 1, $1'$ and $1''$.  

Models of neutrino mixing based on $S_4$ have in fact been studied \cite{Mohapatra:2003tw,Ma:2005pd,Hagedorn:2006ug,Cai:2006mf,Bazzocchi:2008ej,Ishimori:2008fi,Bazzocchi:2009pv,Bazzocchi:2009da,Altarelli:2009gn,Ding:2009iy,Dutta:2009ij,Dutta:2009bj,Meloni:2009cz,Morisi:2010rk,Adulpravitchai:2010na,Hagedorn:2010th,Toorop:2010yh,Ishimori:2010xk,Ding:2010pc,Patel:2010hr}. The group of the permutations of 4 objects $S_4$ has 24 elements and 5 equivalence classes (the character table is given in Tab.~\ref{tcarsS4} that correspond to 5 inequivalent irreducible representations, two singlets, one doublet, two triplets: $1_1$, $1_2$, $2$, $3_1$ and $3_2$). Note that the squares of the dimensions of all these representations add up to 24. 

\begin{vchtable}[h!]
\vchcaption{Characters of $S_4$}
\label{tcarsS4}
\begin{tabular}{@{}cccccc@{}}
\hline 
&&&&&\\[-3mm]
\textbf{Class} & {\boldmath$\chi(1_1)$} & {\boldmath$\chi(1_2)$} &{\boldmath$\chi(2)$}&{\boldmath$\chi(3_1)$}&{\boldmath$\chi(3_2)$}\\[1mm]
\hline 
&&&&&\\[-3mm]
$C_1$ & 1 & 1& 2& 3& 3\\[1mm]
$C_2$ & 1 & 1& 2& -1& -1\\[1mm]
$C_3$ & 1 & -1& 0& 1& -1\\[1mm]
$C_4$ & 1 & 1& -1& 0& 0\\[1mm]
$C_5$ & 1 & -1& 0&-1& 1\\[1mm]
\hline
\end{tabular}
\end{vchtable}

For models of TB mixing, one starts from the $S_4$ presentation $A^3=B^4=(BA^2)^2=1$ and identifies, up to a similarity transformation, $B^2=S$ and $A=T$, where $S$ and $T$ are given in eqs.~(\ref{trep}) and (\ref{ta4}). In this presentation one obtains a realisation of the 3-dimensional representation of $S_4$ where the matrices $S$ and $A_{23}$ in eq.~(\ref{inv}), that leave invariant the TB form of $m_{\nu}$ in eq.~(\ref{gl21}), as well as the matrix $T$ in eq.~(\ref{ta4}), of invariance for $m_e^\dagger m_e$, all explicitly appear \cite{Bazzocchi:2009pv}. In $S_4$ the $1'$ and $1''$ of $A_4$ are collected in a doublet. When the VEV of the doublet flavon is aligned along the $G_S$ preserving direction the resulting couplings are 2-3 symmetric as needed. In $A_4$ the 2-3 symmetry is only achieved if the $1'$ and $1''$ VEV's are identical (which is the $S_4$ prediction). As discussed in Ref. \cite{Bazzocchi:2009pv}, in the leptonic sector the main difference between $A_4$ and $S_4$ is that, while in the typical versions of $A_4$  the most general neutrino mass matrix depends on 2 complex parameters (related to the couplings of the singlet and triplet flavons),  in $S_4$ it depends on 3 complex parameters  (because the doublet is present in addition to singlet and triplet flavons). 

An interesting deformation of the TB mixing pattern arises from the series of groups $\Delta(6 n^2)$ that generalize the permutation group $S_4$, isomorphic to $\Delta(24)$. Indeed the LO mixing pattern induced by $S_4$ is completely determined once the residual symmetries $G_e$ and $G_\nu$ in the charged lepton sector and in the neutrino sector are specified and the TB mixing corresponds to the choice $G_e=Z_3$, $G_\nu=Z_2\times Z_2$. By adopting as flavour group $\Delta(6 n^2)$ $(n=4,8)$, the same choice of residual symmetries leads instead to $U=U_{TB}V$, with $V$ given in eq. (\ref{exact13}), $|\xi'/\alpha|=\tan(\pi/3n)$ and no CKM-like CP violation \cite{Toorop:2011jn,deAdelhartToorop:2011re}. For $n=4,8$, the mixing pattern is close to the experimental data. Concrete models based on $\Delta(96)$ can be found in ref. \cite{Ding:2012xx}. The analysis in Refs.~\cite{Toorop:2011jn,deAdelhartToorop:2011re} accounts only the cases when the residual symmetry in the neutrino sector is $G_\nu=Z_2\times Z_2$; in Refs.~\cite{Hernandez:2012ra,Ge:2011qn,Ge:2011ih}, a more general study has been presented, where the residual symmetry is $G_\nu=Z_2$, while the second $Z_2$ component arises accidentally.

Other flavour groups have been considered for models of TB mixing. Some of them include $S_4$ as a subgroup,  like $PSL_2(7)$ (the smallest group with complex triplet representations) \cite{Luhn:2007yr, King:2009mk, King:2009tj}, while others, like 
 $\Delta(27)$ (which is a discrete subgroup of $SU(3)$)  \cite{deMedeirosVarzielas:2006fc,Luhn:2007uq,Ma:2007wu,Grimus:2008tt,Bazzocchi:2009qg} or $Z_7\rtimes Z_3$ \cite{Luhn:2007sy},  have no direct relation to $S_4$ \cite{King:2009db}. 
 
A different approach to TB mixing has been proposed and developed in different versions by S. King and collaborators  over the last few years \cite{King:2005bj, King:2006me, deMedeirosVarzielas:2005ax, deMedeirosVarzielas:2005qg, King:2009db}. 
The starting point is the decomposition of the neutrino mass matrix given in eqs.~(\ref{1k1}) and (\ref{4k1}) corresponding to exact TB mixing in the diagonal charged
lepton basis:
\begin{equation}
\label{mLL} 
m_\nu= m_1\Phi_1 \Phi_1^T + m_2\Phi_2 \Phi_2^T + m_3\Phi_3 \Phi_3^T
\end{equation}
where $\Phi_1^T=\frac{1}{\sqrt{6}}(2,-1,-1)$,
$\Phi_2^T=\frac{1}{\sqrt{3}}(1,1,1)$, $\Phi_3^T=\frac{1}{\sqrt{2}}(0,-1,1)$,
are the respective columns of $U_{TB}$ and $m_i$ are the neutrino mass eigenvalues.
Such decomposition is purely kinematical and does not possess any dynamical or symmetry
content. In the King models the idea is that the three columns of $U_{TB}$
$\Phi_i$ are promoted to flavon fields whose VEVs break the family symmetry, with the
particular vacuum alignments along the directions
$\Phi_i$. Eq. (\ref{mLL}) directly arises in the see-saw mechanism, $m_\nu=m_D^T M^{-1} m_D$, written in the diagonal RH neutrino mass basis,  $M={\rm diag}(M_1, M_2, M_3)$ when the Dirac
mass matrix is given by $m_D^T=(v_1 \Phi_1,v_2 \Phi_2,v_3 \Phi_3)$, where $v_i$ are mass parameters describing the size of the VEVs.
In this way, to each RH neutrino eigenvalue $M_i$, a particular light neutrino mass $m_i$ is associated. In the case of a strong neutrino hierarchy this idea can be combined with the framework of ``Sequential Dominance'', where the lightest RH neutrino, with its symmetry properties fixes the heaviest light neutrino and so on. For no pronounced hierarchy the correspondence between $M_i$ and $m_i$ can still hold and one talks of ``Form Dominance'' \cite{Chen:2009um}.
In these models
the underlying family symmetry of the Lagrangian $G_f$ is
completely broken by the combined action of the $\Phi_i$ VEV's, and the flavour symmetry of the neutrino mass
matrix emerges entirely as an accidental residual
symmetry of the quadratic form of eq.~(\ref{mLL}) \cite{King:2009db}. The symmetry $G_f$ plays a less direct role and the name ``Indirect Models'' is used by the authors.

An alternative context in which the TB pattern has been implemented is the Holographic Composite Higgs Models: theories in extra dimensions that provide a weakly coupled description of certain 4 dimension composite Higgs models. Neutrinos, as the other fermions and the gauge bosons, represent the elementary sector, while the SM Higgs arise from the composite Higgs sector. The connection between the two sectors is provided through symmetry-defined couplings. In Refs.~\cite{Hagedorn:2011un,Hagedorn:2011pw}, it has been shown that the TB mixing can arise in this context thanks to a non-Abelian discrete symmetry, $S_4$, $A_5$, $\Delta(96)$ and $\Delta(384)$, acting in both the elementary and composite sectors, broken into certain non-trivial subgroups of the original symmetry.

%
%%%%%%%%%%%%%%%%%%%%%%%%%%%  8.  Lepton flavour violation  %%%%%%%%%%%%%%%%%%%
%

\section{Constraints from lepton flavour violating processes}
\label{sec:LFVandLepto}

As we have discussed in the previous sections the relatively large value of $\theta_{13}$ introduces a marked departure from the TB limit, while the values of $\theta_{12}$ and $\theta_{23}$ are very close to it. One challenge for flavour models in the lepton sector is to produce in a natural way a relatively large correction to $\theta_{13}$ without affecting too much the other mixing angles. Another challenge arises from the existing stringent bounds on lepton flavour violating processes. In particular, we refer to the recent improved MEG result \cite{Adam:2011ch} on the $\mu \rightarrow e \gamma$ branching ratio, $Br(\mu \rightarrow e \gamma) \lesssim 2.4\times10^{-12}$ at $95\%$ C.L. and to other similar processes like $\tau \rightarrow (e~\rm{or}~ \mu)  \gamma$. One expects that lepton flavour-violating processes may have a large discriminating power in assessing the relative merits of the different models proposed for neutrino mixing. In fact, one must pay attention that the large corrective terms introduced to shift $\theta_{13}$ from the TB value could appear in the non-diagonal elements of the charged lepton and s-lepton mass matrices (in a basis where all kinetic terms are canonical) and could induce a too large $\mu \to e \gamma$ branching ratio \cite{Feruglio:2008ht,Ishimori:2008au,Feruglio:2009iu,Feruglio:2009hu,Hagedorn:2009df,Chakrabortty:2012vp}. This problem has been discussed in detail in ref. \cite{Altarelli:2012bn} within the simple CMSSM framework (Constrained MSSM). While this GUT-constrained version of supersymmetry is rather marginal after the results of the LHC searches, more so if the Higgs mass is confirmed to lay around $m_H=125$ GeV, still we think that it can be used for our purposes in the present context. 

\begin{figure}[h!]
\centering
\subfigure[{\bf \boldmath Typical $A_4$}: $\tan\beta=2$ and $m_0=200\Gev$]
   {\includegraphics[width=7.3cm]{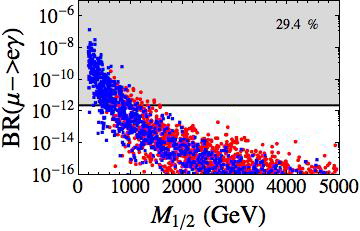}}
\subfigure[{\bf \boldmath Special $A_4$}: $\tan\beta=2$ and $m_0=200\Gev$]
   {\includegraphics[width=7.3cm]{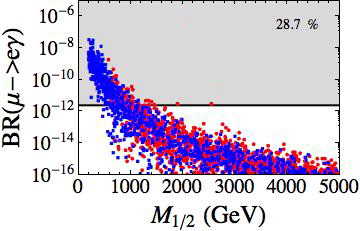}}
 \caption{Scatter plots of $BR(\mu\to e \gamma)$ as a function of $M_{1/2}$, for $\tan\beta=2$, and $m_0= 200$ GeV \cite{Altarelli:2012bn}. The parameters $\xi$ (for typical $A_4$) and $\xi'$ (for special $A_4$)  are chosen in order to maximize the success rate of each model ($\xi=0.075$, $|\xi'|= 0.183$). For Blue (Red) points the lightest supersymmetric mass (LSP) is the lightest neutralino (stau). The percentage in each plot refers to the number of Blue points that satisfy the MEG bound over the total number of points. The horizontal line shows the current MEG bound. For larger values of $\tan\beta$ the success rate decreases, while increases for larger values of $m_0$, being more difficult for the stau to be the LSP.}
\label{fig:MEGA4Wein}
\end{figure}

The results derived in ref.  \cite{Altarelli:2012bn} (see \ref{fig:MEGA4Wein}(a)) show that the typical $A_4$ models are well suited to satisfy the MEG experimental bound, as the non diagonal charged (s-)lepton matrix elements needed to best approximate the mixing angles are particularly small in these models. In fact, the size of the dangerous off diagonal terms is driven by the value of the parameter $\xi$ whose optimal value in the procedure of sect. 4 was found to be $\xi \sim 0.07$. This value corresponds to a modest score in terms of success rate for reproducing the mixing angles in a scanning of the parameter space ( $\sim18\%$ ), but is sufficiently small to maintain the off diagonal charged (s-)lepton mass matrix elements within affordable limits given the bounds on lepton flavour violating processes. A comparable score is also achieved by the models of the Lin type (see see \ref{fig:MEGA4Wein}(b)), because the (on the average) larger correction to the mixing angle $\theta_{13}$ actually arises from the neutrino sector in these models, while the corrections from the charged lepton sector are naturally kept at a smaller level.

In conclusion, when the fit to the mixing angles and the bounds on lepton flavour violating processes are combined, the typical $A_4$ models are rather weak on the mixing angles but, as discussed in detail in Ref.~\cite{Altarelli:2012bn}, are better suited to cope with the bounds on lepton flavour violating processes.  The special $A_4$ models of the Lin type offer the best overall performance to the data as they are far better on the mixing angles and comparable, although at a lower level, on lepton flavour violating processes.  As for the regions of the CMSSM parameter space that are indicated by our analysis the preference is for small $\tan{\beta}$ and large SUSY masses (at least one out of $m_0$ and $m_{1/2}$ must be above 1 TeV).  As a consequence it appears impossible, in these models, at least within the CMSSM framework, to satisfy the MEG bound and, at the same time, to reproduce the muon $g-2$ discrepancy.

%
%%%%%%%%%%%%%%%%%%%%%%%%%%%  10.  Conclusion  %%%%%%%%%%%%%%%%%%%
%
\section{Conclusion}
\label{sec:Conclusions}

The recent rather precise measurements of $\theta_{13}$ make our present knowledge of the neutrino mixing matrix, except for the CP violating phases, sufficiently complete to considerably restrict the class of models that can reproduce the data. In spite of this process the range of possibilities remains unfortunately quite wide. On the one extreme, the rather large value measured for $\theta_{13}$, close to the old CHOOZ bound, has validated the prediction of models based on anarchy \cite{Hall:1999sn,deGouvea:2003xe}, i.e. no symmetry in the leptonic sector, only chance, so that this possibility remains valid, as discussed, for example, in ref. \cite{deGouvea:2012ac}.  Anarchy can be formulated in a $SU(5) \bigotimes U(1)$ context by taking different Froggatt-Nielsen\cite{Froggatt:1978nt} charges only for the $SU(5)$ tenplets (for example $10\sim (3,2,0)$, where 3 is the charge of the first generation, 2 of the second, zero of the third) while no charge differences appear in the $\bar 5$ ($\bar 5\sim (0,0,0)$). Anarchy can be mitigated by assuming that it  only holds in the 2-3 sector with the advantage that the first generation masses and the angle $\theta_{13}$ are naturally small (see also the recent revisiting in ref.\cite{Buchmuller:2011tm}). In models with See-Saw, one can also play with the charges for the right-handed SU(5) singlet neutrinos. If, for example, one takes $1\sim(1, -1, 0)$, together with $\bar 5\sim (2,0,0)$,  it is then possible to get a normal hierarchy model with $\theta_{13}$ small and also with $r = \Delta m^2_{solar}/\Delta m^2_{atm}$ naturally small (see, for example, Ref.~\cite{Altarelli:2002sg}). In summary anarchy and its variants, all based on chance, offer a rather economical class of models that are among those encouraged by the new $\theta_{13}$ result. On the other extreme, stimulated by the fact that the data suggest some special mixing patterns as good first approximations, in particular TB mixing, models based on discrete flavour symmetries, like $A_4$ or $S_4$, have been proposed and widely studied. 

In $A_4$ models, the $A_4$ symmetry is broken down to two different subgroups in the charged lepton sector and in the neutrino sector, and 
the mixing matrix arises from the mismatch between the two different residual symmetries. The breaking can be realized in a natural way through the specific vacuum alignments of a set of scalar  flavons. 
There are many variants of models where TB mixing is indeed derived at leading order (in particular with or without see-saw) with different detailed predictions for the spectrum of neutrino masses and for deviations from the TB values of the mixing angles. The starting LO approximation is completely fixed (no chance), but the NLO corrections introduce a number of undetermined parameters. 
In general at NLO the different mixing angles receive corrections of the same order of magnitude, which are constrained to be small due to the experimental results which are  close to the TB values. Indeed the small experimental error on $\theta_{12}$, with a central value that is close to the value predicted by TB mixing, suggests that the NLO corrections should be of order of few percent, at most.
The recent data on $\theta_{13}$ and the MEG new upper bound on the LFV process $\mu \to e \gamma$ impose a reappraisal of these models \cite{Altarelli:2012bn}. In particular, the relatively large value of $\theta_{13}$ introduces a marked departure from the TB limit, while the values of $\theta_{12}$ and $\theta_{23}$ are very close to it. The challenge is to produce in a natural way a relatively large correction to $\theta_{13}$ without affecting too much the other mixing angles. But one must pay attention that these larger corrective terms introduced to shift $\theta_{13}$ from the TB value could appear in the non-diagonal elements of the charged lepton (and s-lepton) mass matrix and could induce a too large $\mu \to e \gamma$ branching ratio. 

As a result of a detailed analysis \cite{Altarelli:2012bn} we find that, for reproducing the mixing angles, the Lin type $A_4$ models have the best performance, as expected, but the typical $A_4$  models can also accommodate the data with a reasonable probability. As for lepton flavor violating processes, the problem has been studied by adopting the simple CMSSM framework. While this over constrained version of supersymmetry is rather marginal after the results of the LHC searches, more so if the Higgs mass really is around $m_H=125$ GeV, still we think it can be used here for indicative purposes.  The typical $A_4$ models turn out to be the best suited to satisfy the MEG experimental bound, as the non diagonal charged lepton matrix elements needed to reproduce the mixing angles are rather small. A slightly worse score, but still rather good, is achieved by the models of the Lin type, where the main corrections to the mixing angles arise from the neutrino sector. When the fit to the mixing angles and the bounds on LFV processes are combined, the $A_4$ models emerge well from our analysis and in particular those of the Lin type are remarkably successful in the lepton sector.  As for the regions of the CMSSM parameter space that are indicated by our analysis the preference is for small $\tan{\beta}$ and large SUSY masses (at least one out of $m_0$ and $m_{1/2}$ must be above 1 TeV).  As a consequence it appears impossible, at least within the CMSSM model, to satisfy the MEG bound and, at the same time, to reproduce the muon $g-2$ discrepancy.

It is remarkable that neutrino and  quark mixings have such a different qualitative pattern.  An obvious question is whether some additional indication for discrete flavour groups can be obtained by considering the extension of the models to the quark sector, perhaps in a Grand Unified context. The answer appears to be that, while the quark masses and mixings can indeed be reproduced in models where TB (or GR or BM) mixing is realized in the leptonic sector through the action of discrete groups, there are no specific additional hints in favor of discrete groups that come from the quark sector \cite{Altarelli:2010gt}. 

Finally, one could have imagined that neutrinos would bring a decisive boost towards the formulation of a comprehensive understanding of fermion masses and mixings. In reality it is frustrating that no real illumination was sparked on the problem of flavour. We can reproduce the observations in many different ways, in a wide range of models that goes from anarchy to discrete flavour symmetries, but we have not yet been able to single out a unique and convincing baseline for the understanding of fermion masses and mixings. In spite of many interesting ideas and the formulation of many elegant models the mysteries of the flavour structure of the three generations of fermions have not been much unveiled.

%%%%%%%%%%%%%%%%%%%%%%%%%%%%%%%%%%%%%%%%%%%%%%%%%%%%%%%%%%%%%%%%
% Acknowledgements
%%%%%%%%%%%%%%%%%%%%%%%%%%%%%%%%%%%%%%%%%%%%%%%%%%%%%%%%%%%%%%%%
\section*{Acknowledgements}

We recognize that this work has been partly supported by the Italian Ministero dell'Uni\-ver\-si\-t\`a e della Ricerca Scientifica, under the COFIN program (PRIN 2008), by the European Commission, under the networks ``Heptools'', ``Quest for Unification'', ``LHCPHENONET'' and European Union FP7  ITN INVISIBLES (Marie Curie Actions, PITN- GA-2011- 289442)  and contracts MRTN-CT-2006-035505 and  PITN-GA-2009-237920 (UNILHC), and by the Te\-ch\-ni\-sche Universit\"at M\"unchen -- Institute for Advanced Study, funded by the German Excellence Initiative.

%%%%%%%%%%%%%%%%%%%%%%%%%%%%%%%%%%%%%%%%%%%%%%%%%%%%%%%%%%%%%%%%
%  Bibliografy
%%%%%%%%%%%%%%%%%%%%%%%%%%%%%%%%%%%%%%%%%%%%%%%%%%%%%%%%%%%%%%%%

%\bibliography{biblio}{}

\begin{thebibliography}{[100]}

\bibitem{Altarelli:2004za}% article
 \textsc{G.~Altarelli} and  \textsc{F.~Feruglio}\iffalse {Models of Neutrino
  Masses and Mixings}\fi,
 \jr{New J. Phys.} \textbf{6}, 106 (2004).


\bibitem{Mohapatra:2006gs}% article
 \textsc{R.\,N. Mohapatra} and  \textsc{A.\,Y. Smirnov}\iffalse {Neutrino Mass
  and New Physics}\fi,
 \jr{Ann. Rev. Nucl. Part. Sci.} \textbf{56}, 569--628 (2006).


\bibitem{Grimus:2006nb}% article
 \textsc{W.~Grimus}\iffalse {Neutrino physics: Models for neutrino masses and
  lepton mixing}\fi,
 \jr{PoS} \textbf{P2GC}, 001 (2006).


\bibitem{GonzalezGarcia:2007ib}% article
 \textsc{M.\,C. Gonzalez-Garcia} and  \textsc{M.~Maltoni}\iffalse
  {Phenomenology with Massive Neutrinos}\fi,
 \jr{Phys. Rept.} \textbf{460}, 1--129 (2008).


\bibitem{Altarelli:2009wt}% article
 \textsc{G.~Altarelli}\iffalse {Status of Neutrino Masses and Mixing in
  2009}\fi,
 \jr{Nuovo Cim.} \textbf{C32N5-6}, 91--102 (2009).


\bibitem{Altarelli:2010fk}% article
 \textsc{G.~Altarelli}\iffalse {Status of Neutrino Masses and Mixing in
  2010}\fi,
 \jr{PoS} \textbf{HRM${\bf {\rm S}^2}$010}, 022 (2010).


\bibitem{Fogli:2012ua}% article
 \textsc{G.~Fogli},  \textsc{E.~Lisi},  \textsc{A.~Marrone},
  \textsc{D.~Montanino},  \textsc{A.~Palazzo} \etal{}\iffalse {Global Analysis
  of Neutrino Masses, Mixings and Phases: Entering the Era of Leptonic CP
  Violation Searches}\fi, 
  \jr{arXiv:} 1205.5254.


\bibitem{Tortola:2012te}% article
 \textsc{M.~Tortola},  \textsc{J.~Valle},  and  \textsc{D.~Vanegas}\iffalse
  {Global Status of Neutrino Oscillation Parameters After Recent Reactor
  Measurements}\fi,
  \jr{arXiv:} 1205.4018.


\bibitem{Abe:2011sj}% article
 \textsc{K.~Abe} \etal{}\iffalse {Indication of Electron Neutrino Appearance
  from an Accelerator-Produced Off-Axis Muon Neutrino Beam}\fi,
 \jr{Phys. Rev. Lett.} \textbf{107}, 041801 (2011).


\bibitem{Adamson:2011qu}% article
 \textsc{P.~Adamson} \etal{}\iffalse {Improved search for muon-neutrino to
  electron-neutrino oscillations in MINOS}\fi,
 \jr{Phys. Rev. Lett.} \textbf{107}, 181802 (2011).


\bibitem{Abe:2011fz}% article
 \textsc{Y.~Abe} \etal{}\iffalse {Indication for the Disappearance of Reactor
  Electron Antineutrinos in the Double Chooz Experiment}\fi,
  \jr{arXiv:} 1112.6353.


\bibitem{An:2012eh}% article
 \textsc{F.\,P. An} \etal{}\iffalse {Observation of Electron-Antineutrino
  Disappearance at Daya Bay}\fi,
  \jr{arXiv:} 1203.1669.


\bibitem{Ahn:2012nd}% article
 \textsc{J.\,K. Ahn} \etal{}\iffalse {Observation of Reactor Electron
  Antineutrino Disappearance in the Reno Experiment}\fi,
  \jr{arXiv:} 1204.0626.


\bibitem{Altarelli:2010gt}% article
 \textsc{G.~Altarelli} and  \textsc{F.~Feruglio}\iffalse {Discrete Flavor
  Symmetries and Models of Neutrino Mixing}\fi,
 \jr{Rev. Mod. Phys.} \textbf{82}, 2701--2729 (2010).


\bibitem{Ishimori:2010au}% article
 \textsc{H.~Ishimori} \etal{}\iffalse {Non-Abelian Discrete Symmetries in
  Particle Physics}\fi,
 \jr{Prog. Theor. Phys. Suppl.} \textbf{183}, 1--163 (2010).


\bibitem{Ludl:2010bj}% article
 \textsc{P.\,O. Ludl}\iffalse {On the Finite Subgroups of U(3) of Order Smaller
  Than 512}\fi,
 \jr{J. Phys.} \textbf{A43}, 395204 (2010).


\bibitem{Grimus:2010ak}% article
 \textsc{W.~Grimus} and  \textsc{P.\,O. Ludl}\iffalse {Principal Series of
  Finite Subgroups of $SU(3)$}\fi,
 \jr{J. Phys.} \textbf{A43}, 445209 (2010).


\bibitem{Parattu:2010cy}% article
 \textsc{K.\,M. Parattu} and  \textsc{A.~Wingerter}\iffalse {Tribimaximal
  Mixing from Small Groups}\fi,
 \jr{Phys. Rev.} \textbf{D84}, 013011 (2011).


\bibitem{Grimus:2011fk}% article
 \textsc{W.~Grimus} and  \textsc{P.\,O. Ludl}\iffalse {Finite Flavour Groups of
  Fermions}\fi,
  \jr{arXiv:} 1110.6376.


\bibitem{Harrison:2002er}% article
 \textsc{P.\,F. Harrison},  \textsc{D.\,H. Perkins},  and  \textsc{W.\,G.
  Scott}\iffalse {Tri-Bimaximal Mixing and the Neutrino Oscillation Data}\fi,
 \jr{Phys. Lett.} \textbf{B530}, 167 (2002).


\bibitem{Harrison:2002kp}% article
 \textsc{P.\,F. Harrison} and  \textsc{W.\,G. Scott}\iffalse {Symmetries and
  Generalisations of Tri-Bimaximal Neutrino Mixing}\fi,
 \jr{Phys. Lett.} \textbf{B535}, 163--169 (2002).


\bibitem{Xing:2002sw}% article
 \textsc{Z.\,z. Xing}\iffalse {Nearly Tri-Bimaximal Neutrino Mixing and CP
  Violation}\fi,
 \jr{Phys. Lett.} \textbf{B533}, 85--93 (2002).


\bibitem{Harrison:2002et}% article
 \textsc{P.\,F. Harrison} and  \textsc{W.\,G. Scott}\iffalse {Mu - Tau
  Reflection Symmetry in Lepton Mixing and Neutrino Oscillations}\fi,
 \jr{Phys. Lett.} \textbf{B547}, 219--228 (2002).


\bibitem{Harrison:2003aw}% article
 \textsc{P.\,F. Harrison} and  \textsc{W.\,G. Scott}\iffalse {Permutation
  Symmetry, Tri-Bimaximal Neutrino Mixing and the $S^3$ Group Characters}\fi,
 \jr{Phys. Lett.} \textbf{B557}, 76 (2003).


\bibitem{Kajiyama:2007gx}% article
 \textsc{Y.~Kajiyama},  \textsc{M.~Raidal},  and  \textsc{A.~Strumia}\iffalse
  {The Golden Ratio Prediction for the Solar Neutrino Mixing}\fi,
 \jr{Phys. Rev.} \textbf{D76}, 117301 (2007).


\bibitem{Everett:2008et}% article
 \textsc{L.\,L. Everett} and  \textsc{A.\,J. Stuart}\iffalse {Icosahedral (A5)
  Family Symmetry and the Golden Ratio Prediction for Solar Neutrino
  Mixing}\fi,
 \jr{Phys. Rev.} \textbf{D79}, 085005 (2009).


\bibitem{Ding:2011cm}% article
 \textsc{G.\,J. Ding},  \textsc{L.\,L. Everett},  and  \textsc{A.\,J.
  Stuart}\iffalse {Golden Ratio Neutrino Mixing and $A_{5}$ Flavor
  Symmetry}\fi,
 \jr{Nucl. Phys.} \textbf{B857}, 219--253 (2012).


\bibitem{Feruglio:2011qq}% article
 \textsc{F.~Feruglio} and  \textsc{A.~Paris}\iffalse {The Golden Ratio
  Prediction for the Solar Angle from a Natural Model with A5 Flavour
  Symmetry}\fi,
 \jr{JHEP} \textbf{03}, 101 (2011).


\bibitem{Rodejohann:2008ir}% article
 \textsc{W.~Rodejohann}\iffalse {Unified Parametrization for Quark and Lepton
  Mixing Angles}\fi,
 \jr{Phys. Lett.} \textbf{B671}, 267--271 (2009).


\bibitem{Adulpravitchai:2009bg}% article
 \textsc{A.~Adulpravitchai},  \textsc{A.~Blum},  and
  \textsc{W.~Rodejohann}\iffalse {Golden Ratio Prediction for Solar Neutrino
  Mixing}\fi,
 \jr{New J. Phys.} \textbf{11}, 063026 (2009).


\bibitem{Altarelli:2004jb}% article
 \textsc{G.~Altarelli},  \textsc{F.~Feruglio},  and  \textsc{I.~Masina}\iffalse
  {Can Neutrino Mixings Arise from the Charged Lepton Sector?}\fi,
 \jr{Nucl. Phys.} \textbf{B689}, 157--171 (2004).


\bibitem{Raidal:2004iw}% article
 \textsc{M.~Raidal}\iffalse {Relation Between the Neutrino and Quark Mixing
  Angles and Grand Unification}\fi,
 \jr{Phys. Rev. Lett.} \textbf{93}, 161801 (2004).


\bibitem{Minakata:2004xt}% article
 \textsc{H.~Minakata} and  \textsc{A.\,Y. Smirnov}\iffalse {Neutrino Mixing and
  Quark-Lepton Complementarity}\fi,
 \jr{Phys. Rev.} \textbf{D70}, 073009 (2004).


\bibitem{Frampton:2004vw}% article
 \textsc{P.\,H. Frampton} and  \textsc{R.\,N. Mohapatra}\iffalse {Possible
  Gauge Theoretic Origin for Quark-Lepton Complementarity}\fi,
 \jr{JHEP} \textbf{01}, 025 (2005).


\bibitem{Ferrandis:2004vp}% article
 \textsc{J.~Ferrandis} and  \textsc{S.~Pakvasa}\iffalse {Qlc Relation and
  Neutrino Mass Hierarchy}\fi,
 \jr{Phys. Rev.} \textbf{D71}, 033004 (2005).


\bibitem{Kang:2005as}% article
 \textsc{S.\,K. Kang},  \textsc{C.\,S. Kim},  and  \textsc{J.~Lee}\iffalse
  {Quark-Lepton Complementarity with Renormalization Effects Through Threshold
  Corrections}\fi,
 \jr{Phys. Lett.} \textbf{B619}, 129--135 (2005).


\bibitem{Li:2005ir}% article
 \textsc{N.~Li} and  \textsc{B.\,Q. Ma}\iffalse {Unified Parametrization of
  Quark and Lepton Mixing Matrices}\fi,
 \jr{Phys. Rev.} \textbf{D71}, 097301 (2005).


\bibitem{Cheung:2005gq}% article
 \textsc{K.~Cheung},  \textsc{S.\,K. Kang},  \textsc{C.\,S. Kim},  and
  \textsc{J.~Lee}\iffalse {Lepton Flavor Violation as a Probe of Quark-Lepton
  Unification}\fi,
 \jr{Phys. Rev.} \textbf{D72}, 036003 (2005).


\bibitem{Xing:2005ur}% article
 \textsc{Z.\,z. Xing}\iffalse {Nontrivial Correlation Between the Ckm and Mns
  Matrices}\fi,
 \jr{Phys. Lett.} \textbf{B618}, 141--149 (2005).


\bibitem{Datta:2005ci}% article
 \textsc{A.~Datta},  \textsc{L.~Everett},  and  \textsc{P.~Ramond}\iffalse
  {Cabibbo Haze in Lepton Mixing}\fi,
 \jr{Phys. Lett.} \textbf{B620}, 42--51 (2005).


\bibitem{Antusch:2005ca}% article
 \textsc{S.~Antusch},  \textsc{S.\,F. King},  and  \textsc{R.\,N.
  Mohapatra}\iffalse {Quark Lepton Complementarity in Unified Theories}\fi,
 \jr{Phys. Lett.} \textbf{B618}, 150--161 (2005).


\bibitem{Lindner:2005pk}% article
 \textsc{M.~Lindner},  \textsc{M.\,A. Schmidt},  and  \textsc{A.\,Y.
  Smirnov}\iffalse {Screening of Dirac Flavor Structure in the Seesaw and
  Neutrino Mixing}\fi,
 \jr{JHEP} \textbf{07}, 048 (2005).


\bibitem{Minakata:2005rf}% article
 \textsc{H.~Minakata}\iffalse {Quark-Lepton Complementarity: a Review}\fi,
 \jr{arXiv:} hep-ph/0505262.


\bibitem{Ohlsson:2005js}% article
 \textsc{T.~Ohlsson}\iffalse {Bimaximal Fermion Mixing from the Quark and
  Leptonic Mixing Matrices}\fi,
 \jr{Phys. Lett.} \textbf{B622}, 159--164 (2005).


\bibitem{King:2005bj}% article
 \textsc{S.\,F. King}\iffalse {Predicting Neutrino Parameters from S$O(3)$
  Family Symmetry and Quark-Lepton Unification}\fi,
 \jr{JHEP} \textbf{08}, 105 (2005).


\bibitem{Dighe:2006zk}% article
 \textsc{A.~Dighe},  \textsc{S.~Goswami},  and  \textsc{P.~Roy}\iffalse
  {Quark-Lepton Complementarity with Quasidegenerate Majorana Neutrinos}\fi,
 \jr{Phys. Rev.} \textbf{D73}, 071301 (2006).


\bibitem{Chauhan:2006im}% article
 \textsc{B.\,C. Chauhan},  \textsc{M.~Picariello},  \textsc{J.~Pulido},  and
  \textsc{E.~Torrente-Lujan}\iffalse {Quark-Lepton Complementarity, Neutrino
  and Standard Model Data Predict $(\theta_{13}^{PMNS}=9^{+1}_{-2})^\circ$}\fi,
 \jr{Eur. Phys. J.} \textbf{C50}, 573--578 (2007).


\bibitem{Hochmuth:2006xn}% article
 \textsc{K.\,A. Hochmuth} and  \textsc{W.~Rodejohann}\iffalse {Low and High
  Energy Phenomenology of Quark-Lepton Complementarity Scenarios}\fi,
 \jr{Phys. Rev.} \textbf{D75}, 073001 (2007).


\bibitem{Schmidt:2006rb}% article
 \textsc{M.\,A. Schmidt} and  \textsc{A.\,Y. Smirnov}\iffalse {Quark Lepton
  Complementarity and Renormalization Group Effects}\fi,
 \jr{Phys. Rev.} \textbf{D74}, 113003 (2006).


\bibitem{Plentinger:2006nb}% article
 \textsc{F.~Plentinger},  \textsc{G.~Seidl},  and  \textsc{W.~Winter}\iffalse
  {Systematic Parameter Space Search of Extended Quark-Lepton
  Complementarity}\fi,
 \jr{Nucl. Phys.} \textbf{B791}, 60--92 (2008).


\bibitem{Plentinger:2007px}% article
 \textsc{F.~Plentinger},  \textsc{G.~Seidl},  and  \textsc{W.~Winter}\iffalse
  {The Seesaw Mechanism in Quark-Lepton Complementarity}\fi,
 \jr{Phys. Rev.} \textbf{D76}, 113003 (2007).


\bibitem{Altarelli:2009gn}% article
 \textsc{G.~Altarelli},  \textsc{F.~Feruglio},  and  \textsc{L.~Merlo}\iffalse
  {Revisiting Bimaximal Neutrino Mixing in a Model with $S_4$ Discrete
  Symmetry}\fi,
 \jr{JHEP} \textbf{05}, 020 (2009).


\bibitem{Toorop:2010yh}% article
 \textsc{R.~de~Adelhart~Toorop},  \textsc{F.~Bazzocchi},  and
  \textsc{L.~Merlo}\iffalse {The Interplay Between GUT and Flavour Symmetries
  in a Pati-Salam $\times\, S_4$ Model}\fi,
 \jr{JHEP} \textbf{08}, 001 (2010).


\bibitem{Patel:2010hr}% article
 \textsc{K.\,M. Patel}\iffalse {An $SO(10)\times S_4$ Model of Quark-Lepton
  Complementarity}\fi,
 \jr{Phys. Lett.} \textbf{B695}, 225--230 (2011).


\bibitem{Meloni:2011fx}% article
 \textsc{D.~Meloni}\iffalse {Bimaximal mixing and large theta13 in a SUSY SU(5)
  model based on S4}\fi,
 \jr{JHEP} \textbf{10}, 010 (2011).


\bibitem{Shimizu:2010pg}% article
 \textsc{Y.~Shimizu} and  \textsc{R.~Takahashi}\iffalse {Deviations from
  Tri-Bimaximality and Quark-Lepton Complementarity}\fi,
 \jr{Europhys.Lett.} \textbf{93}, 61001 (2011).


\bibitem{Ahn:2011yj}% article
 \textsc{Y.\,H. Ahn},  \textsc{H.\,Y. Cheng},  and  \textsc{S.~Oh}\iffalse
  {Quark-Lepton Complementarity and Tribimaximal Neutrino Mixing from Discrete
  Symmetry}\fi,
 \jr{Phys. Rev.} \textbf{D83}, 076012 (2011).


\bibitem{Altarelli:2012bn}% article
 \textsc{G.~Altarelli},  \textsc{F.~Feruglio},  \textsc{L.~Merlo},  and
  \textsc{E.~Stamou}\iffalse {Discrete Flavour Groups, Thet$A_{1}$3 and Lepton
  Flavour Violation}\fi,
  \jr{arXiv:} 1205.4670.


\bibitem{Lin:2009bw}% article
 \textsc{Y.~Lin}\iffalse {Tri-Bimaximal Neutrino Mixing from $A_{4}$ and
  $\theta_{13} \sim \theta_{C}$}\fi,
 \jr{Nucl. Phys.} \textbf{B824}, 95--110 (2010).


\bibitem{Fukuyama:1997ky}% article
 \textsc{T.~Fukuyama} and  \textsc{H.~Nishiura}\iffalse {Mass Matrix of
  Majorana Neutrinos}\fi,
  \jr{arXiv:} hep-ph/9702253.


\bibitem{Mohapatra:1998ka}% article
 \textsc{R.\,N. Mohapatra} and  \textsc{S.~Nussinov}\iffalse {Bimaximal
  Neutrino Mixing and Neutrino Mass Matrix}\fi,
 \jr{Phys. Rev.} \textbf{D60}, 013002 (1999).


\bibitem{Ma:2001mr}% article
 \textsc{E.~Ma} and  \textsc{M.~Raidal}\iffalse {Neutrino Mass, Muon Anomalous
  Magnetic Moment, and Lepton Flavor Nonconservation}\fi,
 \jr{Phys. Rev. Lett.} \textbf{87}, 011802 (2001).


\bibitem{Lam:2001fb}% article
 \textsc{C.\,S. Lam}\iffalse {A 2-3 Symmetry in Neutrino Oscillations}\fi,
 \jr{Phys. Lett.} \textbf{B507}, 214--218 (2001).


\bibitem{Kitabayashi:2002jd}% article
 \textsc{T.~Kitabayashi} and  \textsc{M.~Yasue}\iffalse {S(2L) Permutation
  Symmetry for Left-Handed Mu and Tau Families and Neutrino Oscillations in an
  $SU(3)$L $\times$ U(1)N Gauge Model}\fi,
 \jr{Phys. Rev.} \textbf{D67}, 015006 (2003).


\bibitem{Ghosal:2003mq}% article
 \textsc{A.~Ghosal}\iffalse {An $SU(2)$L $\times$ U(1)Y Model with Reflection
  Symmetry in View of Recent Neutrino Experimental Result}\fi,
  \jr{arXiv:} hep-ph/0304090.


\bibitem{Grimus:2003vx}% article
 \textsc{W.~Grimus} and  \textsc{L.~Lavoura}\iffalse {Maximal Atmospheric
  Neutrino Mixing and the Small Ratio of Muon to Tau Mass}\fi,
 \jr{J. Phys.} \textbf{G30}, 73--82 (2004).


\bibitem{Koide:2003rx}% article
 \textsc{Y.~Koide}\iffalse {Universal Texture of Quark and Lepton Mass Matrices
  with an Extended Flavor 2 <--> 3 Symmetry}\fi,
 \jr{Phys. Rev.} \textbf{D69}, 093001 (2004).


\bibitem{deGouvea:2004gr}% article
 \textsc{A.~de~Gouvea}\iffalse {Deviation of Atmospheric Mixing from Maximal
  and Structure in the Leptonic Flavor Sector}\fi,
 \jr{Phys. Rev.} \textbf{D69}, 093007 (2004).


\bibitem{Grimus:2004cc}% article
 \textsc{W.~Grimus} \etal{}\iffalse {Non-Vanishing U(E3) and Cos(2 Theta(23))
  from a Broken $Z_2$ Symmetry}\fi,
 \jr{Nucl. Phys.} \textbf{B713}, 151--172 (2005).


\bibitem{Mohapatra:2004hta}% article
 \textsc{R.\,N. Mohapatra} and  \textsc{S.~Nasri}\iffalse {Leptogenesis and Mu
  - Tau Symmetry}\fi,
 \jr{Phys. Rev.} \textbf{D71}, 033001 (2005).


\bibitem{Mohapatra:2005ra}% article
 \textsc{R.\,N. Mohapatra},  \textsc{S.~Nasri},  and  \textsc{H.\,B.
  Yu}\iffalse {Leptogenesis, \Mu-\Tau Symmetry and \Thet$A_{13}$}\fi,
 \jr{Phys. Lett.} \textbf{B615}, 231--239 (2005).


\bibitem{Kitabayashi:2005fc}% article
 \textsc{T.~Kitabayashi} and  \textsc{M.~Yasue}\iffalse {Mu - Tau Symmetry and
  Maximal CP Violation}\fi,
 \jr{Phys. Lett.} \textbf{B621}, 133--138 (2005).


\bibitem{Mohapatra:2005wk}% article
 \textsc{R.\,N. Mohapatra},  \textsc{S.~Nasri},  and  \textsc{H.\,B.
  Yu}\iffalse {Seesaw Right Handed Neutrino as the Sterile Neutrino for
  Lsnd}\fi,
 \jr{Phys. Rev.} \textbf{D72}, 033007 (2005).


\bibitem{Mohapatra:2005yu}% article
 \textsc{R.\,N. Mohapatra} and  \textsc{W.~Rodejohann}\iffalse {Broken Mu-Tau
  Symmetry and Leptonic CP Violation}\fi,
 \jr{Phys. Rev.} \textbf{D72}, 053001 (2005).


\bibitem{Ahn:2006nu}% article
 \textsc{Y.\,H. Ahn},  \textsc{S.\,K. Kang},  \textsc{C.\,S. Kim},  and
  \textsc{J.~Lee}\iffalse {Phased Breaking of Mu - Tau Symmetry and
  Leptogenesis}\fi,
 \jr{Phys. Rev.} \textbf{D73}, 093005 (2006).


\bibitem{Ge:2010js}% article
 \textsc{S.\,F. Ge},  \textsc{H.\,J. He},  and  \textsc{F.\,R. Yin}\iffalse
  {Common Origin of Soft Mu-Tau and CP Breaking in Neutrino Seesaw and the
  Origin of Matter}\fi,
 \jr{JCAP} \textbf{1005}, 017 (2010).


\bibitem{Hagedorn:2010mq}% article
 \textsc{C.~Hagedorn} and  \textsc{R.~Ziegler}\iffalse {Mu-Tau Symmetry and
  Charged Lepton Mass Hierarchy in a Supersymmetric D4 Model}\fi,
 \jr{Phys. Rev.} \textbf{D82}, 053011 (2010).


\bibitem{Low:2003dz}% article
 \textsc{C.\,I. Low} and  \textsc{R.\,R. Volkas}\iffalse {Tri-Bimaximal Mixing,
  Discrete Family Symmetries, and a Conjecture Connecting the Quark and Lepton
  Mixing Matrices}\fi,
 \jr{Phys. Rev.} \textbf{D68}, 033007 (2003).


\bibitem{Antusch:2003kp}% article
 \textsc{S.~Antusch},  \textsc{J.~Kersten},  \textsc{M.~Lindner},  and
  \textsc{M.~Ratz}\iffalse {Running Neutrino Masses, Mixings and CP Phases:
  Analytical Results and Phenomenological Consequences}\fi,
 \jr{Nucl. Phys.} \textbf{B674}, 401--433 (2003).


\bibitem{Antusch:2005gp}% article
 \textsc{S.~Antusch},  \textsc{J.~Kersten},  \textsc{M.~Lindner},
  \textsc{M.~Ratz},  and  \textsc{M.\,A. Schmidt}\iffalse {Running Neutrino
  Mass Parameters in See-Saw Scenarios}\fi,
 \jr{JHEP} \textbf{03}, 024 (2005).


\bibitem{Mei:2005qp}% article
 \textsc{J.\,w. Mei}\iffalse {Running Neutrino Masses, Leptonic Mixing Angles
  and Cp- Violating Phases: from M(Z) to Lambda(Gut)}\fi,
 \jr{Phys. Rev.} \textbf{D71}, 073012 (2005).


\bibitem{Ellis:2005dr}% article
 \textsc{J.\,R. Ellis},  \textsc{A.~Hektor},  \textsc{M.~Kadastik},
  \textsc{K.~Kannike},  and  \textsc{M.~Raidal}\iffalse {Running of Low-Energy
  Neutrino Masses, Mixing Angles and CP Violation}\fi,
 \jr{Phys. Lett.} \textbf{B631}, 32--41 (2005).


\bibitem{Dighe:2007ksa}% article
 \textsc{A.~Dighe},  \textsc{S.~Goswami},  and  \textsc{P.~Roy}\iffalse
  {Radiatively Broken Symmetries of Nonhierarchical Neutrinos}\fi,
 \jr{Phys. Rev.} \textbf{D76}, 096005 (2007).


\bibitem{Lin:2009sq}% article
 \textsc{Y.~Lin},  \textsc{L.~Merlo},  and  \textsc{A.~Paris}\iffalse {Running
  Effects on Lepton Mixing Angles in Flavour Models with Type I Seesaw}\fi,
 \jr{Nucl. Phys.} \textbf{B835}, 238--261 (2010).


\bibitem{Feruglio:2007hi}% article
 \textsc{F.~Feruglio} and  \textsc{Y.~Lin}\iffalse {Fermion Mass Hierarchies
  and Flavour Mixing from a Minimal Discrete Symmetry}\fi,
 \jr{Nucl. Phys.} \textbf{B800}, 77--93 (2008).


\bibitem{Meloni:2012ci}% article
 \textsc{D.~Meloni}\iffalse {$S^3$ as a Flavour Symmetry for Quarks and Leptons
  After the Daya Bay Result on \Theta 13}\fi,
   \jr{arXiv:} 1203.3126.


\bibitem{Cabibbo:1977nk}% article
 \textsc{N.~Cabibbo}\iffalse {Time Reversal Violation in Neutrino
  Oscillation}\fi,
 \jr{Phys. Lett.} \textbf{B72}, 333--335 (1978).


\bibitem{Wolfenstein:1978uw}% article
 \textsc{L.~Wolfenstein}\iffalse {Oscillations among Three Neutrino Types and
  CP Violation}\fi,
 \jr{Phys. Rev.} \textbf{D18}, 958--960 (1978).


\bibitem{Altarelli:2006kg}% article
 \textsc{G.~Altarelli},  \textsc{F.~Feruglio},  and  \textsc{Y.~Lin}\iffalse
  {Tri-Bimaximal Neutrino Mixing from Orbifolding}\fi,
 \jr{Nucl. Phys.} \textbf{B775}, 31--44 (2007).


\bibitem{Altarelli:2005yx}% article
 \textsc{G.~Altarelli} and  \textsc{F.~Feruglio}\iffalse {Tri-Bimaximal
  Neutrino Mixing, $A_4$ and the Modular Symmetry}\fi,
 \jr{Nucl. Phys.} \textbf{B741}, 215--235 (2006).


\bibitem{Ma:2001dn}% article
 \textsc{E.~Ma} and  \textsc{G.~Rajasekaran}\iffalse {Softly Broken $A_{4}$
  Symmetry for Nearly Degenerate Neutrino Masses}\fi,
 \jr{Phys. Rev.} \textbf{D64}, 113012 (2001).


\bibitem{Lavoura:2007dw}% article
 \textsc{L.~Lavoura} and  \textsc{H.~Kuhbock}\iffalse {$A_{4}$ Model for the
  Quark Mass Matrices}\fi,
 \jr{Eur. Phys. J.} \textbf{C55}, 303--308 (2008).


\bibitem{Morisi:2009sc}% article
 \textsc{S.~Morisi} and  \textsc{E.~Peinado}\iffalse {An A4 Model for Lepton
  Masses and Mixings}\fi,
 \jr{Phys. Rev.} \textbf{D80}, 113011 (2009).


\bibitem{Toorop:2010ex}% article
 \textsc{R.~de~Adelhart~Toorop},  \textsc{F.~Bazzocchi},  \textsc{L.~Merlo},
  and  \textsc{A.~Paris}\iffalse {Constraining Flavour Symmetries at the Ew
  Scale I: the A4 Higgs Potential}\fi,
 \jr{JHEP} \textbf{03}, 035 (2011).


\bibitem{Toorop:2010kt}% article
 \textsc{R.~de~Adelhart~Toorop},  \textsc{F.~Bazzocchi},  \textsc{L.~Merlo},
  and  \textsc{A.~Paris}\iffalse {Constraining Flavour Symmetries at the Ew
  Scale Ii: the Fermion Processes}\fi,
 \jr{JHEP} \textbf{03}, 040 (2011).


\bibitem{Cooper:2011rh}% article
 \textsc{I.\,K. Cooper},  \textsc{S.\,F. King},  and  \textsc{C.~Luhn}\iffalse
  {Renormalisation Group Improved Leptogenesis in Family Symmetry Models}\fi,
 \jr{Nucl. Phys.} \textbf{B859}, 159--176 (2012).


\bibitem{Cooper:2012wf}% article
 \textsc{I.\,K. Cooper},  \textsc{S.\,F. King},  and  \textsc{C.~Luhn}\iffalse
  {A4X$SU(5)$ SUSY GUT of Flavour with Trimaximal Neutrino Mixing}\fi
  \jr{arXiv:} 1203.1324.


\bibitem{Froggatt:1978nt}% article
 \textsc{C.\,D. Froggatt} and  \textsc{H.\,B. Nielsen}\iffalse {Hierarchy of
  Quark Masses, Cabibbo Angles and CP Violation}\fi,
 \jr{Nucl. Phys.} \textbf{B147}, 277 (1979).


\bibitem{Lin:2008aj}% article
 \textsc{Y.~Lin}\iffalse {A Predictive $A_4$ Model, Charged Lepton Hierarchy
  and Tri-Bimaximal Sum Rule}\fi,
 \jr{Nucl. Phys.} \textbf{B813}, 91--105 (2009).


\bibitem{Altarelli:2009kr}% article
 \textsc{G.~Altarelli} and  \textsc{D.~Meloni}\iffalse {A Simplest A4 Model for
  Tri-Bimaximal Neutrino Mixing}\fi,
 \jr{J. Phys.} \textbf{G36}, 085005 (2009).


\bibitem{Bazzocchi:2009da}% article
 \textsc{F.~Bazzocchi},  \textsc{L.~Merlo},  and  \textsc{S.~Morisi}\iffalse
  {Phenomenological Consequences of See-Saw in $S_4$ Based Models}\fi,
 \jr{Phys. Rev.} \textbf{D80}, 053003 (2009).


\bibitem{Dorame:2011eb}% article
 \textsc{L.~Dorame},  \textsc{D.~Meloni},  \textsc{S.~Morisi},
  \textsc{E.~Peinado},  and  \textsc{J.\,W.\,F. Valle}\iffalse {Constraining
  Neutrinoless Double Beta Decay}\fi,
  \jr{Nucl.Phys.} \textbf{B861}, 259-270 (2011).


\bibitem{Altarelli:2005yp}% article
 \textsc{G.~Altarelli} and  \textsc{F.~Feruglio}\iffalse {Tri-Bimaximal
  Neutrino Mixing from Discrete Symmetry in Extra Dimensions}\fi,
 \jr{Nucl. Phys.} \textbf{B720}, 64--88 (2005).


\bibitem{King:2007pr}% article
 \textsc{S.\,F. King}\iffalse {Parametrizing the Lepton Mixing Matrix in Terms
  of Deviations from Tri-Bimaximal Mixing}\fi,
 \jr{Phys. Lett.} \textbf{B659}, 244--251 (2008).


\bibitem{Hernandez:2012ra}% article
 \textsc{D.~Hernandez} and  \textsc{A.\,Y. Smirnov}\iffalse {Lepton Mixing and
  Discrete Symmetries}\fi,
  \jr{arXiv:} 1204.0445.
  
  
\bibitem{Varzielas:2010mp}% article
 \textsc{I.~de~Medeiros~Varzielas} and  \textsc{L.~Merlo}\iffalse {Ultraviolet
  Completion of Flavour Models}\fi,
 \jr{JHEP} \textbf{02}, 062 (2011).


\bibitem{King:2011zj}% article
 \textsc{S.\,F. King} and  \textsc{C.~Luhn}\iffalse {Trimaximal Neutrino Mixing
  from Vacuum Alignment in $A_4$ and $S_4$ Models}\fi,
 \jr{JHEP} \textbf{09}, 042 (2011).


\bibitem{Xing:2011at}% article
 \textsc{Z.\,Z. Xing}\iffalse {The T2K Indication of Relatively Large
  Thet$A_{1}$3 and a Natural Perturbation to the Democratic Neutrino Mixing
  Pattern}\fi,
 \jr{Chin. Phys.} \textbf{C36}, 101--105 (2012).


\bibitem{Zheng:2011uz}% article
 \textsc{Y.\,j. Zheng} and  \textsc{B.\,Q. Ma}\iffalse {Re-Evaluation of
  Neutrino Mixing Pattern According to Latest T2K Result}\fi,
 \jr{Eur. Phys. J. Plus} \textbf{127}, 7 (2012).


\bibitem{Ma:2011yi}% article
 \textsc{E.~Ma} and  \textsc{D.~Wegman}\iffalse {Nonzero $\theta_{13}$ for
  Neutrino Mixing in the Context of $A_{4}$ Symmetry}\fi,
 \jr{Phys. Rev. Lett.} \textbf{107}, 061803 (2011).


\bibitem{Zhou:2011nu}% article
 \textsc{S.~Zhou}\iffalse {Relatively Large Theta13 and Nearly Maximal Theta23
  from the Approximate $S^3$ Symmetry of Lepton Mass Matrices}\fi,
 \jr{Phys. Lett.} \textbf{B704}, 291--295 (2011).


\bibitem{Araki:2011wn}% article
 \textsc{T.~Araki}\iffalse {Getting at Large Thet$A_{1}$3 with Almost Maximal
  Thet$A_{2}$3 from Tri-Bimaximal Mixing}\fi,
 \jr{Phys. Rev.} \textbf{D84}, 037301 (2011).


\bibitem{Haba:2011nv}% article
 \textsc{N.~Haba} and  \textsc{R.~Takahashi}\iffalse {Predictions via Large
  Theta13 from Cascades}\fi,
 \jr{Phys. Lett.} \textbf{B702}, 388--393 (2011).


\bibitem{Morisi:2011pm}% article
 \textsc{S.~Morisi},  \textsc{K.\,M. Patel},  and  \textsc{E.~Peinado}\iffalse
  {Model for T2K Indication with Maximal Atmospheric Angle and Tri-Maximal
  Solar Angle}\fi,
 \jr{Phys. Rev.} \textbf{D84}, 053002 (2011).


\bibitem{Chao:2011sp}% article
 \textsc{W.~Chao} and  \textsc{Y.\,j. Zheng}\iffalse {Relatively Large Theta13
  from Modification to the Tri- Bimaximal, Bimaximal and Democratic Neutrino
  Mixing Matrices}\fi,
   \jr{arXiv:} 1107.0738.


\bibitem{Zhang:2011aw}% article
 \textsc{H.~Zhang} and  \textsc{S.~Zhou}\iffalse {Radiative Corrections and
  Explicit Perturbations to the Tetra-Maximal Neutrino Mixing with Large
  Thet$A_{1}$3}\fi,
 \jr{Phys. Lett.} \textbf{B704}, 296--302 (2011).


\bibitem{Dev:2011bd}% article
 \textsc{S.~Dev},  \textsc{S.~Gupta},  and  \textsc{R.\,R. Gautam}\iffalse
  {Parametrizing the Lepton Mixing Matrix in Terms of Charged Lepton
  Corrections}\fi,
 \jr{Phys. Lett.} \textbf{B704}, 527--533 (2011).


\bibitem{Chu:2011jg}% article
 \textsc{X.~Chu},  \textsc{M.~Dhen},  and  \textsc{T.~Hambye}\iffalse
  {Relations among Neutrino Observables in the Light of a Large Thet$A_{1}$3
  Angle}\fi,
 \jr{JHEP} \textbf{11}, 106 (2011).


\bibitem{BhupalDev:2011gi}% article
 \textsc{P.\,S. Bhupal~Dev},  \textsc{R.\,N. Mohapatra},  and
  \textsc{M.~Severson}\iffalse {Neutrino Mixings in SO(10) with Type II Seesaw
  and Thet$A_{13}$}\fi,
 \jr{Phys. Rev.} \textbf{D84}, 053005 (2011).


\bibitem{Toorop:2011jn}% article
 \textsc{R.\,d.\,A. Toorop},  \textsc{F.~Feruglio},  and
  \textsc{C.~Hagedorn}\iffalse {Discrete Flavour Symmetries in Light of
  T2K}\fi,
 \jr{Phys. Lett.} \textbf{B703}, 447--451 (2011).


\bibitem{Antusch:2011qg}% article
 \textsc{S.~Antusch} and  \textsc{V.~Maurer}\iffalse {Large Neutrino Mixing
  Angle \Thet$A_{13}$$^{$Mns} and Quark- Lepton Mass Ratios in Unified Flavour
  Models}\fi,
 \jr{Phys. Rev.} \textbf{D84}, 117301 (2011).


\bibitem{Rodejohann:2011uz}% article
 \textsc{W.~Rodejohann},  \textsc{H.~Zhang},  and  \textsc{S.~Zhou}\iffalse
  {Systematic Search for Successful Lepton Mixing Patterns with Nonzero
  Thet$A_{1}$3}\fi,
 \jr{Nucl. Phys.} \textbf{B855}, 592--607 (2012).


\bibitem{Ahn:2011if}% article
 \textsc{Y.\,H. Ahn},  \textsc{H.\,Y. Cheng},  and  \textsc{S.~Oh}\iffalse {An
  Extension of Tribimaximal Lepton Mixing}\fi,
 \jr{Phys. Rev.} \textbf{D84}, 113007 (2011).


\bibitem{Marzocca:2011dh}% article
 \textsc{D.~Marzocca},  \textsc{S.\,T. Petcov},  \textsc{A.~Romanino},  and
  \textsc{M.~Spinrath}\iffalse {Sizeable \Thet$A_{1}$3 from the Charged Lepton
  Sector in $SU(5)$, (Tri-)Bimaximal Neutrino Mixing and Dirac CP
  Violation}\fi,
 \jr{JHEP} \textbf{11}, 009 (2011).


\bibitem{Ge:2011qn}% article
 \textsc{S.\,F. Ge},  \textsc{D.\,A. Dicus},  and  \textsc{W.\,W.
  Repko}\iffalse {Residual Symmetries for Neutrino Mixing with a Large
  Thet$A_{1}$3 and Nearly Maximal Delt$A_{D}$}\fi,
 \jr{Phys. Rev. Lett.} \textbf{108}, 041801 (2012).


\bibitem{Kumar:2011vf}% article
 \textsc{S.~Kumar}\iffalse {Implications of a Class of Neutrino Mass Matrices
  with Texture Zeros for Non-Zero \Thet$A_{13}$}\fi,
 \jr{Phys. Rev.} \textbf{D84}, 077301 (2011).


\bibitem{Bazzocchi:2011ax}% article
 \textsc{F.~Bazzocchi}\iffalse {Tri-Permuting Mixing Matrix and Predictions for
  Thet$A_{1}$3}\fi,
  \jr{arXiv:} 1108.2497.


\bibitem{Araki:2011qy}% article
 \textsc{T.~Araki} and  \textsc{C.\,Q. Geng}\iffalse {Large \Thet$A_{1}$3 from
  Finite Quantum Corrections in Quasi- Degenerate Neutrino Mass Spectrum}\fi,
 \jr{JHEP} \textbf{09}, 139 (2011).


\bibitem{Antusch:2011ic}% article
 \textsc{S.~Antusch},  \textsc{S.\,F. King},  \textsc{C.~Luhn},  and
  \textsc{M.~Spinrath}\iffalse {Trimaximal Mixing with Predicted \Thet$A_{1}$3
  from a New Type of Constrained Sequential Dominance}\fi,
 \jr{Nucl. Phys.} \textbf{B856}, 328--341 (2012).


\bibitem{Fritzsch:2011qv}% article
 \textsc{H.~Fritzsch},  \textsc{Z.\,z. Xing},  and  \textsc{S.~Zhou}\iffalse
  {Two-Zero Textures of the Majorana Neutrino Mass Matrix and Current
  Experimental Tests}\fi,
 \jr{JHEP} \textbf{09}, 083 (2011).


\bibitem{Rashed:2011zs}% article
 \textsc{A.~Rashed} and  \textsc{A.~Datta}\iffalse {The Charged Lepton Mass
  Matrix and Non-Zero \Thet$A_{13}$ with TeV Scale New Physics}\fi,
 \jr{Phys. Rev.} \textbf{D85}, 035019 (2012).


\bibitem{Ludl:2011vv}% article
 \textsc{P.\,O. Ludl},  \textsc{S.~Morisi},  and  \textsc{E.~Peinado}\iffalse
  {The Reactor Mixing Angle and CP Violation with Two Texture Zeros in the
  Light of T2K}\fi,
 \jr{Nucl. Phys.} \textbf{B857}, 411--423 (2012).


\bibitem{Verma:2011kz}% article
 \textsc{S.~Verma}\iffalse {Non-Zero \Thet$A_{13}$ and Cp-Violation in Inverse
  Neutrino Mass Matrix}\fi,
 \jr{Nucl. Phys.} \textbf{B854}, 340--349 (2012).


\bibitem{Meloni:2011ac}% article
 \textsc{D.~Meloni}\iffalse {Large Thet$A_{1}$3 from a Model with Broken
  L_E-L_Mu-L_Tau Symmetry}\fi,
 \jr{JHEP} \textbf{02}, 090 (2012).


\bibitem{Dev:2011hf}% article
 \textsc{S.~Dev},  \textsc{S.~Gupta},  \textsc{R.\,R. Gautam},  and
  \textsc{L.~Singh}\iffalse {Near Maximal Atmospheric Mixing in Neutrino Mass
  Matrices with Two Vanishing Minors}\fi,
 \jr{Phys. Lett.} \textbf{B706}, 168--176 (2011).


\bibitem{Deepthi:2011sk}% article
 \textsc{K.\,N. Deepthi},  \textsc{S.~Gollu},  and  \textsc{R.~Mohanta}\iffalse
  {Neutrino Mixing Matrices with Relatively Large \Thet$A_{13}$ and with
  Texture One-Zero}\fi,
 \jr{Eur. Phys. J.} \textbf{C72}, 1888 (2012).


\bibitem{Rashed:2011xe}% article
 \textsc{A.~Rashed}\iffalse {Deviation from Tri-Bimaximal Mixing and Large
  Reactor Mixing Angle}\fi,
  \jr{arXiv:} 1111.3072.


\bibitem{deAdelhartToorop:2011re}% article
 \textsc{R.~de~Adelhart~Toorop},  \textsc{F.~Feruglio},  and
  \textsc{C.~Hagedorn}\iffalse {Finite Modular Groups and Lepton Mixing}\fi
  (2011).


\bibitem{deMedeirosVarzielas:2011wx}% article
 \textsc{I.~de~Medeiros~Varzielas}\iffalse {Non-Abelian Family Symmetries in
  Pati-Salam Unification}\fi,
 \jr{JHEP} \textbf{01}, 097 (2012).


\bibitem{Araki:2011zg}% article
 \textsc{T.~Araki} and  \textsc{Y.\,F. Li}\iffalse {Q_6 Flavor Symmetry Model
  for the Extension of the Minimal Standard Model by Three Right-Handed Sterile
  Neutrinos}\fi,
 \jr{Phys. Rev.} \textbf{D85}, 065016 (2012).


\bibitem{Gupta:2011ct}% article
 \textsc{S.~Gupta},  \textsc{A.\,S. Joshipura},  and  \textsc{K.\,M.
  Patel}\iffalse {Minimal Extension of Tri-Bimaximal Mixing and Generalized Z_2
  $\times$ Z_2 Symmetries}\fi,
 \jr{Phys. Rev.} \textbf{D85}, 031903 (2012).


\bibitem{Ding:2012xx}% article
 \textsc{G.\,J. Ding}\iffalse {Tfh Mixing Patterns, Large \Thet$A_{13}$ and
  \Delta(96) Flavor Symmetry}\fi,
  \jr{arXiv:} 1201.3279.


\bibitem{Ishimori:2012gv}% article
 \textsc{H.~Ishimori} and  \textsc{T.~Kobayashi}\iffalse {Lepton Flavor Models
  with Discrete Prediction of Thet$A_{13}$}\fi,
    \jr{arXiv:} 1201.3429.


\bibitem{Dev:2012ns}% article
 \textsc{S.~Dev},  \textsc{R.\,R. Gautam},  and  \textsc{L.~Singh}\iffalse
  {Broken $S^3$ Symmetry in the Neutrino Mass Matrix and Non- Zero
  Thet$A_{13}$}\fi,
 \jr{Phys. Lett.} \textbf{B708}, 284--289 (2012).


\bibitem{Bazzocchi:2012ve}% article
 \textsc{F.~Bazzocchi},  \textsc{S.~Morisi},  \textsc{E.~Peinado},
  \textsc{J.\,W.\,F. Valle},  and  \textsc{A.~Vicente}\iffalse {Flavored
  Bilinear R-Parity Violation}\fi,
  \jr{arXiv:} 1202.1529.


\bibitem{BhupalDev:2012nm}% article
 \textsc{P.\,S. Bhupal~Dev},  \textsc{B.~Dutta},  \textsc{R.\,N. Mohapatra},
  and  \textsc{M.~Severson}\iffalse {\Thet$A_{13}$ and Proton Decay in a
  Minimal SO(10) $\times$ S_4 Model of Flavor}\fi,
  \jr{arXiv:} 1202.4012.


\bibitem{Siyeon:2012zu}% article
 \textsc{K.~Siyeon}\iffalse {Non-Vanishing U_{E3} Under S_3 Symmetry}\fi,
  \jr{arXiv:} 1203.1593.


\bibitem{Wu:2012ri}% article
 \textsc{Y.\,L. Wu}\iffalse {$SU(3)$ Gauge Family Symmetry and Prediction for
  the Lepton- Flavor Mixing and Neutrino Masses with Maximal Spontaneous CP
  Violation}\fi,
    \jr{arXiv:} 1203.2382.


\bibitem{Branco:2012vs}% article
 \textsc{G.\,C. Branco},  \textsc{R.\,G. Felipe},  \textsc{F.\,R. Joaquim},
  and  \textsc{H.~Serodio}\iffalse {Spontaneous Leptonic CP Violation and
  Nonzero \Thet$A_{13}$}\fi,
  \jr{arXiv:} 1203.2646.


\bibitem{Ahn:2012tv}% article
 \textsc{Y.\,H. Ahn} and  \textsc{S.\,K. Kang}\iffalse {Non-Zero $\theta_{13}$
  and CP Violation in a Model with $A_{4}$ Flavor Symmetry}\fi,
  \jr{arXiv:} 1203.4185.


\bibitem{Varzielas:2012ss}% article
 \textsc{I.\,d.\,M. Varzielas} and  \textsc{G.\,G. Ross}\iffalse {Discrete
  Family Symmetry, Higgs Mediators and $\theta_{13}$}\fi,
  \jr{arXiv:} 1203.6636.


\bibitem{Hagedorn:2012pg}% article
 \textsc{C.~Hagedorn} and  \textsc{D.~Meloni}\iffalse {D14 - a Common Origin of
  the Cabibbo Angle and the Lepton Mixing Angle Theta$^l$_13}\fi,
  \jr{arXiv:} 1204.0715.


\bibitem{Hagedorn:2012ut}% article
 \textsc{C.~Hagedorn},  \textsc{S.\,F. King},  and  \textsc{C.~Luhn}\iffalse
  {SUSY S4 $\times$ $SU(5)$ Revisited}\fi,
  \jr{arXiv:} 1205.3114.


\bibitem{Carr:2007qw}% article
 \textsc{P.\,D. Carr} and  \textsc{P.\,H. Frampton}\iffalse {Group Theoretic
  Bases for Tribimaximal Mixing}\fi,
  \jr{arXiv:} hep-ph/0701034.


\bibitem{Feruglio:2007uu}% article
 \textsc{F.~Feruglio},  \textsc{C.~Hagedorn},  \textsc{Y.~Lin},  and
  \textsc{L.~Merlo}\iffalse {Tri-Bimaximal Neutrino Mixing and Quark Masses
  from a Discrete Flavour Symmetry}\fi,
 \jr{Nucl. Phys.} \textbf{B775}, 120--142 (2007).


\bibitem{Chen:2007afa}% article
 \textsc{M.\,C. Chen} and  \textsc{K.\,T. Mahanthappa}\iffalse {Ckm and
  Tri-Bimaximal Mns Matrices in a $SU(5)$ $\times$ (D)T Model}\fi,
 \jr{Phys. Lett.} \textbf{B652}, 34--39 (2007).


\bibitem{Frampton:2007et}% article
 \textsc{P.\,H. Frampton} and  \textsc{T.\,W. Kephart}\iffalse {Flavor Symmetry
  for Quarks and Leptons}\fi,
 \jr{JHEP} \textbf{09}, 110 (2007).


\bibitem{Aranda:2007dp}% article
 \textsc{A.~Aranda}\iffalse {Neutrino Mixing from the Double Tetrahedral Group
  T$^{$\Prime}}\fi,
 \jr{Phys. Rev.} \textbf{D76}, 111301 (2007).


\bibitem{Ding:2008rj}% article
 \textsc{G.\,J. Ding}\iffalse {Fermion Mass Hierarchies and Flavor Mixing from
  T' Symmetry}\fi,
 \jr{Phys. Rev.} \textbf{D78}, 036011 (2008).


\bibitem{Frampton:2009fw}% article
 \textsc{P.\,H. Frampton} and  \textsc{S.~Matsuzaki}\iffalse {T$^{$'}
  Predictions of Pmns and Ckm Angles}\fi,
 \jr{Phys. Lett.} \textbf{B679}, 347--349 (2009).


\bibitem{Chen:2009gf}% article
 \textsc{M.\,C. Chen} and  \textsc{K.~Mahanthappa}\iffalse {Group Theoretical
  Origin of CP Violation}\fi,
 \jr{Phys.Lett.} \textbf{B681}, 444--447 (2009).


\bibitem{Merlo:2011hw}% article
 \textsc{L.~Merlo},  \textsc{S.~Rigolin},  and  \textsc{B.~Zaldivar}\iffalse
  {Flavour Violation in a Supersymmetric $T'$ Model}\fi,
 \jr{JHEP} \textbf{11}, 047 (2011).


\bibitem{Barbieri:1995uv}% article
 \textsc{R.~Barbieri},  \textsc{G.\,R. Dvali},  and  \textsc{L.\,J.
  Hall}\iffalse {Predictions from a U(2) Flavour Symmetry in Supersymmetric
  Theories}\fi,
 \jr{Phys. Lett.} \textbf{B377}, 76--82 (1996).


\bibitem{Barbieri:1996ww}% article
 \textsc{R.~Barbieri},  \textsc{L.\,J. Hall},  \textsc{S.~Raby},  and
  \textsc{A.~Romanino}\iffalse {Unified Theories with U(2) Flavor Symmetry}\fi,
 \jr{Nucl. Phys.} \textbf{B493}, 3--26 (1997).


\bibitem{Barbieri:1997tu}% article
 \textsc{R.~Barbieri},  \textsc{L.\,J. Hall},  and
  \textsc{A.~Romanino}\iffalse {Consequences of a U(2) Flavour Symmetry}\fi,
 \jr{Phys. Lett.} \textbf{B401}, 47--53 (1997).


\bibitem{Altarelli:2008bg}% article
 \textsc{G.~Altarelli},  \textsc{F.~Feruglio},  and
  \textsc{C.~Hagedorn}\iffalse {A SUSY $SU(5)$ Grand Unified Model of
  Tri-Bimaximal Mixing from A4}\fi,
 \jr{JHEP} \textbf{03}, 052--052 (2008).


\bibitem{Kawamura:2000ev}% article
 \textsc{Y.~Kawamura}\iffalse {Triplet-Doublet Splitting, Proton Stability and
  Extra Dimension}\fi,
 \jr{Prog. Theor. Phys.} \textbf{105}, 999--1006 (2001).


\bibitem{Ma:2005tr}% article
 \textsc{E.~Ma}\iffalse {Hiding the Existence of a Family Symmetry in the
  Standard Model}\fi,
 \jr{Mod. Phys. Lett.} \textbf{A20}, 2767--2774 (2005).


\bibitem{Ma:2006sk}% article
 \textsc{E.~Ma},  \textsc{H.~Sawanaka},  and  \textsc{M.~Tanimoto}\iffalse
  {Quark Masses and Mixing with $A_{4}$ Family Symmetry}\fi,
 \jr{Phys. Lett.} \textbf{B641}, 301--304 (2006).


\bibitem{Ma:2006wm}% article
 \textsc{E.~Ma}\iffalse {Suitability of $A_{4}$ as a Family Symmetry in Grand
  Unification}\fi,
 \jr{Mod. Phys. Lett.} \textbf{A21}, 2931--2936 (2006).


\bibitem{Morisi:2007ft}% article
 \textsc{S.~Morisi},  \textsc{M.~Picariello},  and
  \textsc{E.~Torrente-Lujan}\iffalse {A Model for Fermion Masses and Lepton
  Mixing in SO(10) $\times$ A4}\fi,
 \jr{Phys. Rev.} \textbf{D75}, 075015 (2007).


\bibitem{Grimus:2008tm}% article
 \textsc{W.~Grimus} and  \textsc{H.~Kuhbock}\iffalse {Embedding the
  Zee-Wolfenstein Neutrino Mass Matrix in an SO(10) $\times$ A4 GUT
  Scenario}\fi,
 \jr{Phys. Rev.} \textbf{D77}, 055008 (2008).


\bibitem{Ciafaloni:2009ub}% article
 \textsc{P.~Ciafaloni},  \textsc{M.~Picariello},  \textsc{E.~Torrente-Lujan},
  and  \textsc{A.~Urbano}\iffalse {Neutrino Masses and Tribimaximal Mixing in
  Minimal Renormalizable SUSY $SU(5)$ Grand Unified Model with A4 Flavor
  Symmetry}\fi,
 \jr{Phys. Rev.} \textbf{D79}, 116010 (2009).


\bibitem{Bazzocchi:2008rz}% article
 \textsc{F.~Bazzocchi},  \textsc{S.~Morisi},  \textsc{M.~Picariello},  and
  \textsc{E.~Torrente-Lujan}\iffalse {Embedding A4 into $SU(3)$xU(1) Flavor
  Symmetry: Large Neutrino Mixing and Fermion Mass Hierarchy in SO(10) GUT}\fi,
 \jr{J. Phys.} \textbf{G36}, 015002 (2009).


\bibitem{Antusch:2010es}% article
 \textsc{S.~Antusch},  \textsc{S.\,F. King},  and  \textsc{M.~Spinrath}\iffalse
  {Measurable Neutrino Mass Scale in A4 $\times$ $SU(5)$}\fi,
 \jr{Phys. Rev.} \textbf{D83}, 013005 (2011).


\bibitem{Ishimori:2010xk}% article
 \textsc{H.~Ishimori},  \textsc{K.~Saga},  \textsc{Y.~Shimizu},  and
  \textsc{M.~Tanimoto}\iffalse {Tri-Bimaximal Mixing and Cabibbo Angle in $S_4$
  Flavor Model with SUSY}\fi,
 \jr{Phys. Rev.} \textbf{D81}, 115009 (2010).


\bibitem{Hagedorn:2010th}% article
 \textsc{C.~Hagedorn},  \textsc{S.\,F. King},  and  \textsc{C.~Luhn}\iffalse {A
  SUSY GUT of Flavour with $S_4\times SU(5)$ to NLO}\fi,
 \jr{JHEP} \textbf{06}, 048 (2010).


\bibitem{Ding:2010pc}% article
 \textsc{G.\,J. Ding}\iffalse {SUSY Adjoint $SU(5)$ Grand Unified Model with
  $S_4$ Flavor Symmetry}\fi,
 \jr{Nucl. Phys.} \textbf{B846}, 394--428 (2011).


\bibitem{Antusch:2011sx}% article
 \textsc{S.~Antusch},  \textsc{S.\,F. King},  \textsc{C.~Luhn},  and
  \textsc{M.~Spinrath}\iffalse {Right Unitarity Triangles and Tri-Bimaximal
  Mixing from Discrete Symmetries and Unification}\fi,
 \jr{Nucl.Phys.} \textbf{B850}, 477--504 (2011).


\bibitem{Dutta:2009ij}% article
 \textsc{B.~Dutta},  \textsc{Y.~Mimura},  and  \textsc{R.\,N.
  Mohapatra}\iffalse {Origin of Quark-Lepton Flavor in $SO(10)$ with Type II
  Seesaw}\fi,
 \jr{Phys. Rev.} \textbf{D80}, 095021 (2009).


\bibitem{Adulpravitchai:2010na}% article
 \textsc{A.~Adulpravitchai} and  \textsc{M.\,A. Schmidt}\iffalse {Flavored
  Orbifold GUT - an $SO(10)\times S_4$ Model}\fi,
 \jr{JHEP} \textbf{01}, 106 (2011).


\bibitem{King:2009mk}% article
 \textsc{S.\,F. King} and  \textsc{C.~Luhn}\iffalse {A New Family Symmetry for
  SO(10) Guts}\fi,
 \jr{Nucl. Phys.} \textbf{B820}, 269--289 (2009).


\bibitem{King:2009tj}% article
 \textsc{S.\,F. King} and  \textsc{C.~Luhn}\iffalse {A Supersymmetric Grand
  Unified Theory of Flavour with P$SL(2,7)$ $\times$ SO(10)}\fi,
 \jr{Nucl. Phys.} \textbf{B832}, 414--439 (2010).


\bibitem{Bazzocchi:2008sp}% article
 \textsc{F.~Bazzocchi},  \textsc{M.~Frigerio},  and  \textsc{S.~Morisi}\iffalse
  {Fermion Masses and Mixing in Models with SO(10) $\times$ $A_{4}$
  Symmetry}\fi,
 \jr{Phys. Rev.} \textbf{D78}, 116018 (2008).


\bibitem{Altarelli:2010at}% article
 \textsc{G.~Altarelli} and  \textsc{G.~Blankenburg}\iffalse {Different SO(10)
  Paths to Fermion Masses and Mixings}\fi,
 \jr{JHEP} \textbf{03}, 133 (2011).


\bibitem{Joshipura:2011rr}% article
 \textsc{A.\,S. Joshipura} and  \textsc{K.\,M. Patel}\iffalse {Viability of the
  Exact Tri-Bimaximal Mixing at M_{Gut} in SO(10)}\fi,
 \jr{JHEP} \textbf{09}, 137 (2011).


\bibitem{Joshipura:2011nn}% article
 \textsc{A.\,S. Joshipura} and  \textsc{K.\,M. Patel}\iffalse {Fermion Masses
  in SO(10) Models}\fi,
 \jr{Phys. Rev.} \textbf{D83}, 095002 (2011).


\bibitem{Blankenburg:2011vw}% article
 \textsc{G.~Blankenburg} and  \textsc{S.~Morisi}\iffalse {Fermion Masses and
  Mixing with Tri-Bimaximal in SO(10) with Type-I Seesaw}\fi,
 \jr{JHEP} \textbf{01}, 016 (2012).


\bibitem{Burrows:2009pi}% article
 \textsc{T.\,J. Burrows} and  \textsc{S.\,F. King}\iffalse {$A_{4}$ Family
  Symmetry from $SU(5)$ SUSY Guts in 6D}\fi,
 \jr{Nucl. Phys.} \textbf{B835}, 174--196 (2010).


\bibitem{Adulpravitchai:2009id}% article
 \textsc{A.~Adulpravitchai},  \textsc{A.~Blum},  and
  \textsc{M.~Lindner}\iffalse {Non-Abelian Discrete Flavor Symmetries from
  T$^2$/Z_N Orbifolds}\fi,
 \jr{JHEP} \textbf{07}, 053 (2009).


\bibitem{Kobayashi:2006wq}% article
 \textsc{T.~Kobayashi},  \textsc{H.\,P. Nilles},  \textsc{F.~Ploger},
  \textsc{S.~Raby},  and  \textsc{M.~Ratz}\iffalse {Stringy Origin of
  Non-Abelian Discrete Flavor Symmetries}\fi,
 \jr{Nucl. Phys.} \textbf{B768}, 135--156 (2007).


\bibitem{Abe:2009vi}% article
 \textsc{H.~Abe},  \textsc{K.\,S. Choi},  \textsc{T.~Kobayashi},  and
  \textsc{H.~Ohki}\iffalse {Non-Abelian Discrete Flavor Symmetries from
  Magnetized/Intersecting Brane Models}\fi,
 \jr{Nucl. Phys.} \textbf{B820}, 317--333 (2009).


\bibitem{Adulpravitchai:2009kd}% article
 \textsc{A.~Adulpravitchai},  \textsc{A.~Blum},  and
  \textsc{M.~Lindner}\iffalse {Non-Abelian Discrete Groups from the Breaking of
  Continuous Flavor Symmetries}\fi,
 \jr{JHEP} \textbf{09}, 018 (2009).


\bibitem{Berger:2009tt}% article
 \textsc{J.~Berger} and  \textsc{Y.~Grossman}\iffalse {Model of Leptons from
  S$O(3)$ \To $A_{4}$}\fi,
 \jr{JHEP} \textbf{02}, 071 (2010).


\bibitem{Luhn:2011ip}% article
 \textsc{C.~Luhn}\iffalse {Spontaneous Breaking of $SU(3)$ to Finite Family
  Symmetries: a Pedestrian's Approach}\fi,
 \jr{JHEP} \textbf{03}, 108 (2011).


\bibitem{Lam:2008sh}% article
 \textsc{C.\,S. Lam}\iffalse {The Unique Horizontal Symmetry of Leptons}\fi,
 \jr{Phys. Rev.} \textbf{D78}, 073015 (2008).


\bibitem{Grimus:2009pg}% article
 \textsc{W.~Grimus},  \textsc{L.~Lavoura},  and  \textsc{P.\,O. Ludl}\iffalse
  {Is S4 the Horizontal Symmetry of Tri-Bimaximal Lepton Mixing?}\fi,
 \jr{J. Phys.} \textbf{G36}, 115007 (2009).


\bibitem{Mohapatra:2003tw}% article
 \textsc{R.\,N. Mohapatra},  \textsc{M.\,K. Parida},  and
  \textsc{G.~Rajasekaran}\iffalse {High Scale Mixing Unification and Large
  Neutrino Mixing Angles}\fi,
 \jr{Phys. Rev.} \textbf{D69}, 053007 (2004).


\bibitem{Ma:2005pd}% article
 \textsc{E.~Ma}\iffalse {Neutrino Mass Matrix from $S_4$ Symmetry}\fi,
 \jr{Phys. Lett.} \textbf{B632}, 352--356 (2006).


\bibitem{Hagedorn:2006ug}% article
 \textsc{C.~Hagedorn},  \textsc{M.~Lindner},  and  \textsc{R.\,N.
  Mohapatra}\iffalse {$S_4$ Flavor Symmetry and Fermion Masses: Towards a Grand
  Unified Theory of Flavor}\fi,
 \jr{JHEP} \textbf{06}, 042 (2006).


\bibitem{Cai:2006mf}% article
 \textsc{Y.~Cai} and  \textsc{H.\,B. Yu}\iffalse {An $SO(10)$ GUT Model with
  $S_4$ Flavor Symmetry}\fi,
 \jr{Phys. Rev.} \textbf{D74}, 115005 (2006).


\bibitem{Bazzocchi:2008ej}% article
 \textsc{F.~Bazzocchi} and  \textsc{S.~Morisi}\iffalse {$S_4$ as a Natural
  Flavor Symmetry for Lepton Mixing}\fi,
 \jr{Phys. Rev.} \textbf{D80}, 096005 (2009).


\bibitem{Ishimori:2008fi}% article
 \textsc{H.~Ishimori},  \textsc{Y.~Shimizu},  and  \textsc{M.~Tanimoto}\iffalse
  {$S_4$ Flavor Symmetry of Quarks and Leptons in $SU(5)$ GUT}\fi,
 \jr{Prog. Theor. Phys.} \textbf{121}, 769--787 (2009).


\bibitem{Bazzocchi:2009pv}% article
 \textsc{F.~Bazzocchi},  \textsc{L.~Merlo},  and  \textsc{S.~Morisi}\iffalse
  {Fermion Masses and Mixings in a $S_4$-Based Model}\fi,
 \jr{Nucl. Phys.} \textbf{B816}, 204--226 (2009).


\bibitem{Ding:2009iy}% article
 \textsc{G.\,J. Ding}\iffalse {Fermion Masses and Flavor Mixings in a Model
  with $S_4$ Flavor Symmetry}\fi,
 \jr{Nucl. Phys.} \textbf{B827}, 82--111 (2010).


\bibitem{Dutta:2009bj}% article
 \textsc{B.~Dutta},  \textsc{Y.~Mimura},  and  \textsc{R.\,N.
  Mohapatra}\iffalse {An $SO(10)$ Grand Unified Theory of Flavor}\fi,
 \jr{JHEP} \textbf{05}, 034 (2010).


\bibitem{Meloni:2009cz}% article
 \textsc{D.~Meloni}\iffalse {A See-Saw $S_4$ Model for Fermion Masses and
  Mixings}\fi,
 \jr{J. Phys.} \textbf{G37}, 055201 (2010).


\bibitem{Morisi:2010rk}% article
 \textsc{S.~Morisi} and  \textsc{E.~Peinado}\iffalse {An $S_4$ Model for Quarks
  and Leptons with Maximal Atmospheric Angle}\fi,
 \jr{Phys. Rev.} \textbf{D81}, 085015 (2010).


\bibitem{Ge:2011ih}% article
 \textsc{S.\,F. Ge},  \textsc{D.\,A. Dicus},  and  \textsc{W.\,W.
  Repko}\iffalse {Z_2 Symmetry Prediction for the Leptonic Dirac CP Phase}\fi,
 \jr{Phys.Lett.} \textbf{B702}, 220--223 (2011).


\bibitem{Luhn:2007yr}% article
 \textsc{C.~Luhn},  \textsc{S.~Nasri},  and  \textsc{P.~Ramond}\iffalse {Simple
  Finite Non-Abelian Flavor Groups}\fi,
 \jr{J. Math. Phys.} \textbf{48}, 123519 (2007).


\bibitem{deMedeirosVarzielas:2006fc}% article
 \textsc{I.~de~Medeiros~Varzielas},  \textsc{S.\,F. King},  and  \textsc{G.\,G.
  Ross}\iffalse {Neutrino Tri-Bi-Maximal Mixing from a Non-Abelian Discrete
  Family Symmetry}\fi,
 \jr{Phys. Lett.} \textbf{B648}, 201--206 (2007).


\bibitem{Luhn:2007uq}% article
 \textsc{C.~Luhn},  \textsc{S.~Nasri},  and  \textsc{P.~Ramond}\iffalse {The
  Flavor Group Delta(3$ N^2$)}\fi,
 \jr{J. Math. Phys.} \textbf{48}, 073501 (2007).


\bibitem{Ma:2007wu}% article
 \textsc{E.~Ma}\iffalse {Near Tribimaximal Neutrino Mixing with Delta(27)
  Symmetry}\fi,
 \jr{Phys. Lett.} \textbf{B660}, 505--507 (2008).


\bibitem{Grimus:2008tt}% article
 \textsc{W.~Grimus} and  \textsc{L.~Lavoura}\iffalse {A Model for Trimaximal
  Lepton Mixing}\fi,
 \jr{JHEP} \textbf{09}, 106 (2008).


\bibitem{Bazzocchi:2009qg}% article
 \textsc{F.~Bazzocchi} and  \textsc{I.~de~Medeiros~Varzielas}\iffalse
  {Tri-Bi-Maximal Mixing in Viable Family Symmetry Unified Model with Extended
  Seesaw}\fi,
 \jr{Phys. Rev.} \textbf{D79}, 093001 (2009).


\bibitem{Luhn:2007sy}% article
 \textsc{C.~Luhn},  \textsc{S.~Nasri},  and  \textsc{P.~Ramond}\iffalse
  {Tri-Bimaximal Neutrino Mixing and the Family Symmetry Z_7 \Rtimes Z_3}\fi,
 \jr{Phys. Lett.} \textbf{B652}, 27--33 (2007).


\bibitem{King:2009db}% article
 \textsc{S.\,F. King}\iffalse {Neutrino Mass and Flavour Models}\fi,
 \jr{AIP Conf. Proc.} \textbf{1200}, 103--111 (2010).


\bibitem{King:2006me}% article
 \textsc{S.\,F. King} and  \textsc{M.~Malinsky}\iffalse {Towards a Complete
  Theory of Fermion Masses and Mixings with S$O(3)$ Family Symmetry and 5D
  SO(10) Unification}\fi,
 \jr{JHEP} \textbf{11}, 071 (2006).


\bibitem{deMedeirosVarzielas:2005ax}% article
 \textsc{I.~de~Medeiros~Varzielas} and  \textsc{G.\,G. Ross}\iffalse {$SU(3)$
  Family Symmetry and Neutrino Bi-Tri-Maximal Mixing}\fi,
 \jr{Nucl. Phys.} \textbf{B733}, 31--47 (2006).


\bibitem{deMedeirosVarzielas:2005qg}% article
 \textsc{I.~de~Medeiros~Varzielas},  \textsc{S.\,F. King},  and  \textsc{G.\,G.
  Ross}\iffalse {Tri-Bimaximal Neutrino Mixing from Discrete Subgroups of
  $SU(3)$ and S$O(3)$ Family Symmetry}\fi,
 \jr{Phys. Lett.} \textbf{B644}, 153--157 (2007).


\bibitem{Chen:2009um}% article
 \textsc{M.\,C. Chen} and  \textsc{S.\,F. King}\iffalse {A4 See-Saw Models and
  Form Dominance}\fi,
 \jr{JHEP} \textbf{06}, 072 (2009).


\bibitem{Hagedorn:2011un}% article
 \textsc{C.~Hagedorn} and  \textsc{M.~Serone}\iffalse {Leptons in Holographic
  Composite Higgs Models with Non- Abelian Discrete Symmetries}\fi,
 \jr{JHEP} \textbf{10}, 083 (2011).


\bibitem{Hagedorn:2011pw}% article
 \textsc{C.~Hagedorn} and  \textsc{M.~Serone}\iffalse {General Lepton Mixing in
  Holographic Composite Higgs Models}\fi,
 \jr{JHEP} \textbf{02}, 077 (2012).


\bibitem{Adam:2011ch}% article
 \textsc{J.~Adam} \etal{}\iffalse {New Limit on the Lepton-Flavour Violating
  Decay $\mu \to e \gamma$}\fi,
  \jr{arXiv:} 1107.5547.


\bibitem{Feruglio:2008ht}% article
 \textsc{F.~Feruglio},  \textsc{C.~Hagedorn},  \textsc{Y.~Lin},  and
  \textsc{L.~Merlo}\iffalse {Lepton Flavour Violation in Models with $A_4$
  Flavour Symmetry}\fi,
 \jr{Nucl. Phys.} \textbf{B809}, 218--243 (2009).


\bibitem{Ishimori:2008au}% article
 \textsc{H.~Ishimori},  \textsc{T.~Kobayashi},  \textsc{Y.~Omura},  and
  \textsc{M.~Tanimoto}\iffalse {Soft supersymmetry breaking terms from A(4)
  lepton flavor symmetry}\fi,
 \jr{JHEP} \textbf{0812}, 082 (2008).


\bibitem{Feruglio:2009iu}% article
 \textsc{F.~Feruglio},  \textsc{C.~Hagedorn},  and  \textsc{L.~Merlo}\iffalse
  {Vacuum Alignment in SUSY $A_4$ Models}\fi,
 \jr{JHEP} \textbf{03}, 084 (2010).


\bibitem{Feruglio:2009hu}% article
 \textsc{F.~Feruglio},  \textsc{C.~Hagedorn},  \textsc{Y.~Lin},  and
  \textsc{L.~Merlo}\iffalse {Lepton Flavour Violation in a Supersymmetric Model
  with $A_4$ Flavour Symmetry}\fi,
 \jr{Nucl. Phys.} \textbf{B832}, 251--288 (2010).


\bibitem{Hagedorn:2009df}% article
 \textsc{C.~Hagedorn},  \textsc{E.~Molinaro},  and  \textsc{S.\,T.
  Petcov}\iffalse {Charged Lepton Flavour Violating Radiative Decays $\ell_i
  \to \ell_j + \gamma$ in See-Saw Models with $A_{4}$ Symmetry}\fi,
 \jr{JHEP} \textbf{02}, 047 (2010).


 \bibitem{Chakrabortty:2012vp}% article
 \textsc{J.~Chakrabortty}, \textsc{P.~Ghosh} and  \textsc{W.~Rodejohann}\iffalse {Lower Limits on $\mu \to e \gamma$ from new Measurements on $U_{e3}$}\fi,
 \jr{arXiv:} 1204.1000.


\bibitem{Hall:1999sn}% article
 \textsc{L.\,J. Hall},  \textsc{H.~Murayama},  and  \textsc{N.~Weiner}\iffalse
  {Neutrino Mass Anarchy}\fi,
 \jr{Phys. Rev. Lett.} \textbf{84}, 2572--2575 (2000).


\bibitem{deGouvea:2003xe}% article
 \textsc{A.~de~Gouvea} and  \textsc{H.~Murayama}\iffalse {Statistical Test of
  Anarchy}\fi,
 \jr{Phys. Lett.} \textbf{B573}, 94--100 (2003).


\bibitem{deGouvea:2012ac}% article
 \textsc{A.~de~Gouvea} and  \textsc{H.~Murayama}\iffalse {Neutrino Mixing
  Anarchy: Alive and Kicking}\fi
  \jr{arXiv:} 1204.1249.


\bibitem{Buchmuller:2011tm}% article
 \textsc{W.~Buchmuller},  \textsc{V.~Domcke},  and  \textsc{K.~Schmitz}\iffalse
  {Predicting Thet$A_{1}$3 and the Neutrino Mass Scale from Quark Lepton Mass
  Hierarchies}\fi,
 \jr{JHEP} \textbf{03}, 008 (2012).


\bibitem{Altarelli:2002sg}% article
 \textsc{G.~Altarelli},  \textsc{F.~Feruglio},  and  \textsc{I.~Masina}\iffalse
  {Models of Neutrino Masses: Anarchy Versus Hierarchy}\fi,
 \jr{JHEP} \textbf{01}, 035 (2003).


\end{thebibliography}
%\bibliographystyle{fdp}

\providecommand{\WileyBibTextsc}{}
\let\textsc\WileyBibTextsc
\providecommand{\othercit}{}
\providecommand{\jr}[1]{#1}
\providecommand{\etal}{~et~al.}

\end{document}